\documentclass[iop]{emulateapj}%[preprint2]{aastex}
\usepackage{graphicx}

\shorttitle{High spatial resolution Fe\,{\sc xii} spectral
  observations of solar active regions}
\shortauthors{Testa, De Pontieu \& Hansteen}
\usepackage{color} 
\usepackage{amsmath}           
\usepackage{amsfonts}

\def \deg {$^\circ$}
\def \arcsec {\hbox{$^{\prime\prime}$}}

\def \wnth {$w_{\rm nth}$}
\def \vdop {$v_{\rm D}$}

\def \iris   {{\em IRIS}}
\def \sdo   {{\em SDO}}
\def \hinode   {{\em Hinode}}
\def \xrt {{\sc XRT}}
\def \eis  {{\sc EIS}}
\def \aia  {{\sc AIA}}
\def \hrts  {{\sc HRTS}}
\def \sumer  {{\sc SUMER}}

\def \nv     {N\,{\sc v}}
\def \neviii {Ne\,{\sc viii}}
\def \oiv     {O\,{\sc iv}}
\def \ov     {O\,{\sc v}}
\def \oi      {O\,{\sc i}}

\def \fexii   {Fe\,{\sc xii}}
\def \fexiii  {Fe\,{\sc xiii}}

\def \fexvi   {Fe\,{\sc xvi}}

\def \fexviii {Fe\,{\sc xviii}}

\def \s9     {S\,{\sc ix}}

\def \civ    {C\,{\sc iv}}
\def \siiv    {Si\,{\sc iv}}

\def \caii   {Ca\,{\sc ii}}

\def\ion[#1 #2]{#1\,{\sc #2}}
\def\densr[#1 #2]{10$^{#1}$\hskip 1pt{--}\hskip .5pt{10$^{#2}$}\hskip 1.5pt{cm$^{-3}$}}
\def\fl[#1 #2]{{#1}$\pm${#2}}
\def\orb[#1 #2]{{$#1^{#2}$}}
\def\ls[#1 #2]{{$^{#1}${#2}}}
\def\tm[#1 #2 #3]{{$^{#1}${#2}$_{#3}$}}

\newcounter{ion}

\begin{document}

\title{High spatial resolution Fe\,{\sc xii} observations of solar active regions}

\author{Paola Testa$^{1}$}
\email{ptesta@cfa.harvard.edu}
\author{Bart De Pontieu$^{2,3}$}
\author{Viggo Hansteen$^{3}$}

\affil{$^1$ Smithsonian Astrophysical Observatory,60 Garden street, MS
  58, Cambridge, MA 02138, USA} 
\affil{$^2$Lockheed-Martin Solar and Astrophysics Laboratory, 3251 Hanover st., Org. A021S,
Bldg.252, Palo Alto, CA, 94304, USA.}
\affil{$^3$Institute of Theoretical Astrophysics, University of Oslo, P.O. Box 1029, Blindern,
N-0315, Oslo, Norway.}

\begin{abstract}
  We use UV spectral observations of active regions with the
  Interface Region Imaging Spectrograph (\iris) to investigate the
  properties of the coronal \fexii\ 1349.4\AA\ emission at
  unprecedented high spatial resolution ($\sim 0.33$ \arcsec). We
  find that by using appropriate observational strategies (i.e., long
  exposures, lossless compression), \fexii\ emission can be studied
  with \iris\ at high spatial and spectral resolution, at least for
  high density plasma (e.g., post-flare loops, and active region
  moss). 
  We find that upper transition region (moss) \fexii\ emission shows
  very small average Doppler redshifts (\vdop\ $\sim 3$~km~s$^{-1}$),
  as well as modest non-thermal velocities (with an average $\sim
  24$~km~s$^{-1}$, and the peak of the distribution at $\sim
  15$~km~s$^{-1}$).  The observed distribution of Doppler shifts
  appears to be compatible with advanced 3D radiative MHD simulations
  in which impulsive heating is concentrated at the transition region
  footpoints of a hot corona.
  While the non-thermal broadening of \fexii\ 1349.4\AA\ peaks at
  similar values as lower resolution simultaneous \hinode-\eis\
  measurements of \fexii\ 195\AA, \iris\ observations show a
  previously undetected tail of increased non-thermal broadening that
  might be suggestive of the presence of subarcsecond heating
  events. We find that \iris\ and \eis\ non-thermal line broadening
  measurements are affected by instrumental effects that can only be
  removed through careful analysis. 
  Our results also reveal an unexplained discrepancy between observed
  195.1/1349.4\AA\ \fexii\ intensity ratios and those predicted by the
  CHIANTI atomic database.
\end{abstract}


\section{Introduction}
\label{s:intro}

The nature and properties of the heating of the outer atmosphere of
the Sun and solar-like stars to millions of degrees are issues that
remain largely unsolved \citep[e.g., reviews by][]{Klimchuk06,testa15}.
The complex problem of coronal heating is addressed by constraining
models through observational determinations of the plasma conditions
(temperature, density, flows, and their spatial and temporal
distribution throughout the atmosphere) in solar and stellar coronae. 
Solar observations are a particularly powerful tool to investigate the
heating processes, because they provide high spatial and temporal
resolution, for both imaging and spectroscopic data, and significant
progress has been recently made.
Two main candidate heating mechanisms have been studied and modeled in
some detail: dissipation of magnetohydrodynamic (Alfv$\acute{\rm e}$n)
waves \citep[e.g.,][]{vanballegooijen11,vanballegooijen14}, and
dissipation of magnetic stresses in small scale reconnection events
(``nanoflares''), due to random photospheric motions that lead to
braiding of magnetic field lines
\citep[e.g.,][]{parker88,galsgaard96,cargill96,priest02,Gudiksen05a,hansteen15}.  

Observational diagnostics of coronal heating are however often
difficult to achieve because of the typically very small (compared to
current resolution capabilities) spatial and temporal scales of heating
release predicted by most viable heating processes
\citep[e.g.,][]{Klimchuk06,reale14}. 
Furthermore, the immediate plasma response to impulsive heating
events is challenging to detect because of several effects, including
non-equilibrium ionization and the low emission measure causing a very
faint emission from the hot plasma \citep[e.g.,][]{reale14,testa11}. 
In flares the heated plasma is dense and bright, allowing the
  study of the heating processes, but in the non-flaring corona the
detectability of the heated plasma is a significant challenge.
For the non-flaring corona, the study of the transition region
emission provides a powerful alternative approach to the study of
coronal heating as it overcomes several issues present in coronal
observations: the emission of the dense transition region is (a)
bright, (b) confined to a small atmospheric layer (compared to the
large coronal volumes) therefore dramatically reducing the line of
sight overlap of a different structures, (c) very sensitive to heating
events. 
For these reasons, in this paper we mostly focus on this transition
region emission, and in particular on ``moss'', i.e., the upper
transition region (TR) layer at the footpoints of high pressure loops
in active regions, which is very bright in observations sensitive to
$\sim 1$~MK emission
\citep[e.g.,][]{Peres94,Fletcher99,Berger99,depontieu99,Martens00,Brooks09,Tripathi10,winebarger13,testa13,Testa14}.  

In this paper we focus on spectral observations, which provide
particularly useful constraints on coronal heating models. The
spectral line profiles (Doppler shifts, and line broadening) of
coronal and transition region emission diagnose plasma motions (e.g., 
flows and turbulence) that can be compared with expectations from
model. 
For instance, significant non-thermal line broadening (i.e.,
broadening in excess of the thermal and instrumental broadening), is
predicted by several reconnection based models
\citep[e.g.,][]{cargill96,Patsourakos06} and Alfv$\acute{\rm e}$n
wave models \citep[e.g.,][]{vanballegooijen11,asgari-targhi14}. 
Impulsive heating models predict a range of Doppler shifts
\citep[e.g.,][]{Patsourakos06,Hansteen10,taroyan14} depending
on the details of the model (e.g., spatial and temporal distribution
of heating) as discussed in more detail in \S~\ref{s:conclusions}.

Non-thermal line broadening and Doppler shifts in plasma in the solar
outer atmosphere have been studied since the {\em Skylab} era
\citep[e.g.,][]{doschek76,doschek77,doschek81} in a variety of
chromospheric, transition region, and coronal spectral lines.  
In the following decades, the Ultraviolet Spectrometer on the Solar
Maximum Mission \citep{smm}, the NRL High Resolution Telescope and
Spectrograph (\hrts; \citealt{hrts}), the Solar Ultraviolet
Measurements of Emitted Radiation spectrometer (\sumer;
\citealt{sumer}) onboard {\em SOHO}, and the Extreme-ultraviolet
Imaging Spectrometer (\eis; \citealt{Culhane07}) onboard \hinode\
\citep{Kosugi07} have provided spectral observations with improved
spatial and spectral resolution.  
Measurements before \eis\ however mostly focused on transition region
lines formed below $\sim 1$~MK, and/or on quiet sun or coronal holes
\citep[e.g.,][]{athay83,dere84,warren97,chae98,chae98b,akiyama05}.
Early spectroscopic active region observations of emission lines
formed around $\sim 1$~MK in a variety of solar coronal features
  (quiet Sun and active regions, including both on disk and off-limb
  observations) indicated that the non-thermal width (\wnth) was
characterized by relatively small \wnth\ ($\sim 10-25$~km~s$^{-1}$;
e.g., \citealt{cheng79}). Fewer early studies investigated the Doppler
shifts of coronal lines (more uncertain than line broadening due to
their dependence on the accuracy of rest wavelength and of absolute
wavelength calibration) and the results pointed to a range of values
from small blueshifts \citep[e.g.,][]{sandlin77} to small redshifts ($\lesssim
10$~km~s$^{-1}$; \citealt[e.g.,][]{dere82,achour95}), to relatively
large redshifts ($10-30$~km~s$^{-1}$; \citealt{brueckner81}).   
Several recent studies based on \eis\ spectra have helped better
determine the spectral line properties of coronal emission lines in
active regions. \cite{Brooks09} studied Doppler shifts and non-thermal
velocities in active region moss using the \eis\ \fexii\ 195\AA\ line,
and find \wnth\ values in the 15-30~km~s$^{-1}$ range, and Doppler
shifts from $\sim -5$~km~s$^{-1}$ (i.e., blueshifts) to $\sim
10$~km~s$^{-1}$ (redshift). \cite{tripathi12} investigated Doppler
shifts in moss emission and, using a different method than
\cite{Brooks09} for defining a reference wavelength, found for \fexii\
a symmetric distribution ($-10-10$~km~s$^{-1}$) centered around zero
shift. \cite{dadashi12} using yet another calibration method for the
wavelength, finds blueshifts in \fexii\ with typical values of $-10$
to $-5$~km~s$^{-1}$ in moss.

Here we present a first analysis of spectroscopic observations of
\fexii\ emission, obtained at unprecedented spatial resolution of
$\sim 0.33$\arcsec\ with the Interface Region Imaging Spectrograph
(\iris; \citealt{depontieu14}). \iris\ spatial resolution represents
an improvement by a factor of at least 5 ($\sim 16$ \iris\ pixel for
each \eis\ pixel) with respect to previous
spectrographs observing the chromosphere/transition region/corona in
UV/EUV/X-ray spectral wavelengths. Also, as hinted by recent coronal
studies \citep{Warren08loops,testa13,peter13,antolin15}, the collective
behavior in the corona is starting to be resolved at spatial scales of
the order of $\sim 0.3$\arcsec.
\iris\ provides slit-jaw images (SJI)
and high resolution spectra in the UV range, allowing to study the
chromospheric and transition region emission at high resolution.
The presence of chromospheric neutral lines in the \iris\ spectra
allows a much more accurate calibration of the absolute wavelength,
compared with \eis.
The \iris\ FUV spectral range includes a \fexii\ forbidden line at
1349.4\AA\ (3s$^{2}$ 3p$^{3}$ $^4$S$_{3/2}$ - 3s$^{2}$ 3p$^{3}$
$^2$P$_{1/2}$). This coronal line (peak formation temperature around
$\log T$[K]$\sim 6.2$) was observed with {\em Skylab}
\citep[e.g.,][]{doschek77,sandlin77}, together with a stronger
forbidden line of \fexii\ at 1242\AA\ to provide also coronal density
diagnostics \citep[e.g.,]{cook94}. Both lines hower were too weak to be observed
on disk and were generally detected with {\em Skylab} only in long
exposure off-limb spectra \citep{feldman83}, and with very limited
spatial resolution. These coronal forbidden lines were also observed
by \sumer, but very few active region measurements (in the stronger
1242\AA) are available \citep[e.g.,][]{teriaca99,depontieu09}.

\begin{figure*}[!ht]
\centerline{\includegraphics[scale=0.5]{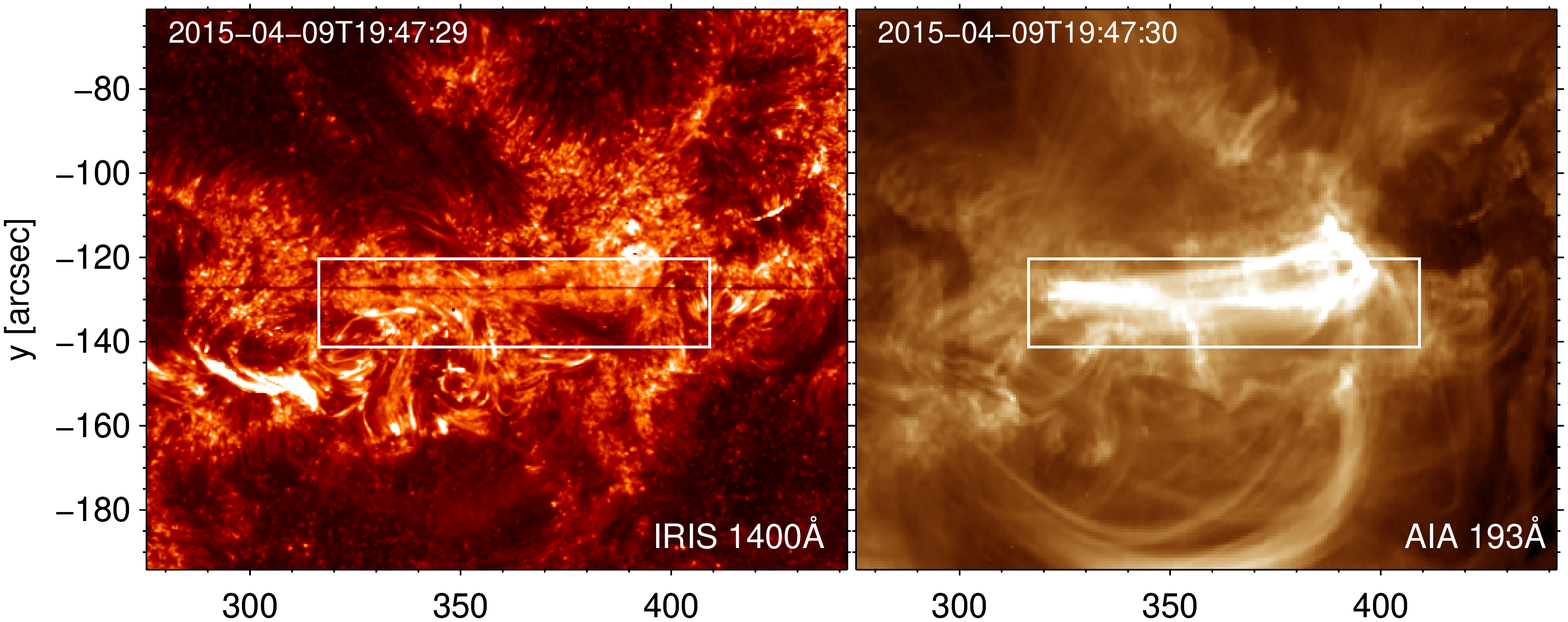}}\vspace{-0.6cm}
\centerline{\includegraphics[scale=0.5]{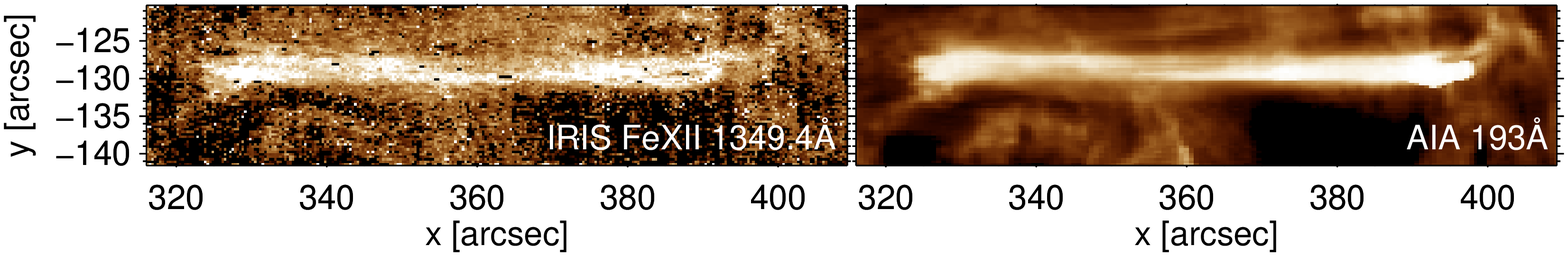}}
\caption{\iris\ and \sdo-\aia\ observations of AR 12320. {\em Top:}
  \iris\ slit-jaw image in the 1400\AA\ passband ({\em left}), and
  \aia\ 193\AA\ image ({\em right}). For a subregion, indicated by the
  box in the \iris\ SJI and \aia\ images, we also show the intensity map in
  the \fexii\ 1349\AA\ line (from Gaussian fit to the \iris\ spectra;
  {\em bottom left}), and the corresponding composite \aia\ 193\AA\
  image (built by selecting, for each horizontal slice of the \iris\
  field of view (f.o.v.) the \aia\ data closest in time; {\em bottom right}).  
  \label{fig:loops}} 
\end{figure*}

In this paper we show that the 1349\AA\ \fexii\ line, though weak, can
be observed with \iris\ at high spatial resolution, provided that long
enough ($\sim 30$s) exposure times and appropriate observing modes are
used (see \S~\ref{s:data} for details). Here we analyze two datasets
with bright targets, post-flare loops and active region moss, for
which the \fexii\ emission can be observed with \iris\ at good
signal-to-noise (see \S~\ref{s:data}).  For the moss dataset we also
present a detailed comparison with coordinated \eis\ observations,
focusing in particular on the \fexii\ 195\AA\ emission.
In section~\ref{s:data} we describe the data selection process and
selected datasets, while in section~\ref{s:results} we address analysis
methods and results. In section~\ref{s:sims} we present the results
of a 3D radiative MHD simulation we use as an aid for the
interpretation of the observed spectral properties of the \fexii\ moss
emission. 
We discuss our findings and draw our conclusions in section~\ref{s:conclusions}.

\section{Data Selection and Reduction}
\label{s:data} 
The goal of this work is to analyze \fexii\ spectral properties for
the first time at the high spatial resolution of \iris\ observations. 
The \fexii\ 1349.4\AA\ forbidden line is characterized by very modest
typical emission levels compared to the much stronger allowed \fexii\
transitions at EUV wavelengths; e.g., the \eis\ \fexii\ 195.12\AA\ line,
which we will analyze in this paper as well, has emission $\gtrsim
2$ orders of magnitude higher than the \iris\ 1349.4\AA\ line,
according to predictions based on the CHIANTI atomic database
\citep{chianti,chianti7,chianti8}.   
Inspection  of \iris\ observations quickly shows that exposure times
of the order of 30s are needed to obtain high enough signal-to-noise
ratio (SNR) to analyze the \fexii\ emission at the highest \iris\
spatial resolution. 

In this paper we present results for two \iris\ active region
datasets, which have relatively bright \fexii\ emission.
The first dataset, shown in Figure~\ref{fig:loops}, is part of a
series of observations we tailored to study the \fexii\ emission in
active regions. 
For the selected sequence \iris\ observed AR 12320 (2015-04-09
19:24-20:31UT), with 64-step dense rasters (in ``dense'' rasters the
raster step is 0.35\arcsec), 30s exposure times, $2 \times 2$ spatial
and spectral binning, lossless compression (\iris\ OBSID 3810112091),
and a roll angle of -90\deg\ (i.e., the slit was oriented in the E-W
direction). Slit-jaw images (SJI) were obtained at each raster
position, alternating all four passbands (1330\AA, 1400\AA, 2796\AA,
2832\AA) therefore yielding a cadence of $\sim 125$s in each
passband. In this dataset both the FUV and NUV spectra have $\Delta
\lambda \sim 0.025$\AA, and their field of view (f.o.v.) is $\sim
21$\arcsec$\times 175$\arcsec.  The center of the \iris\ f.o.v.\ is at
354\arcsec, -132\arcsec.

\begin{figure*}[!ht]
\centerline{\includegraphics[scale=0.67]{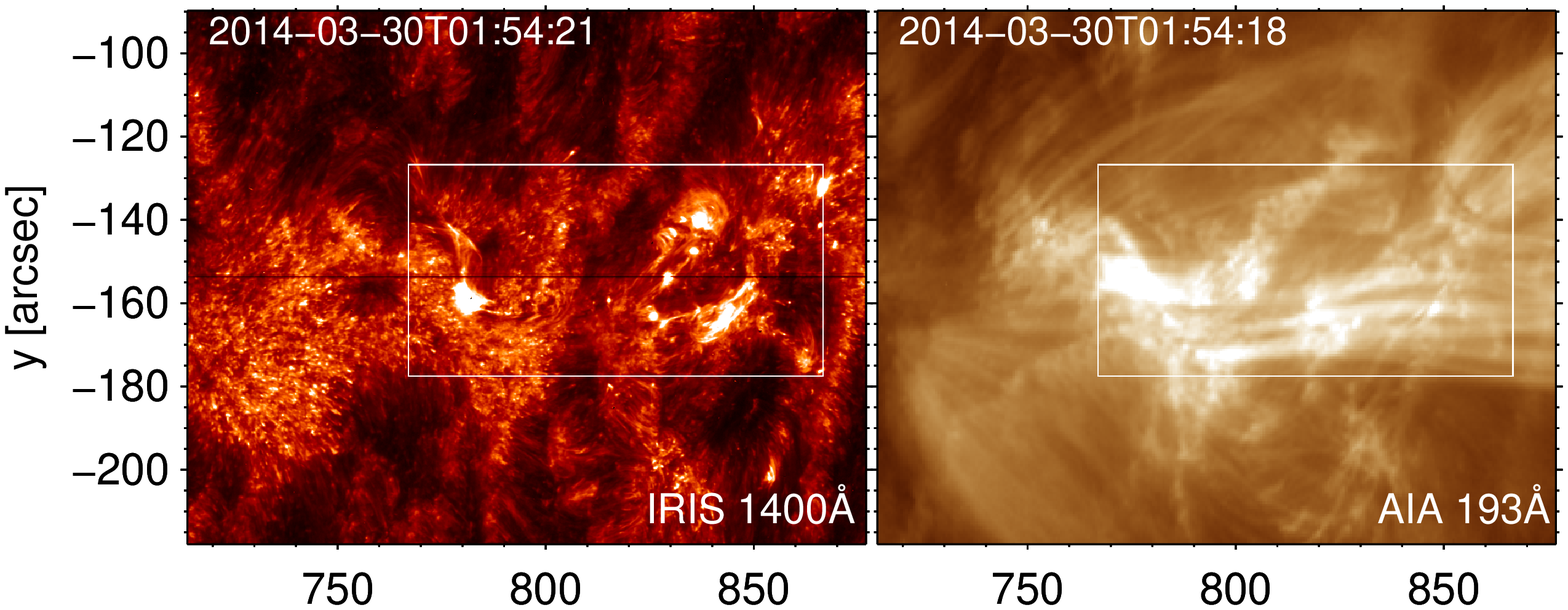}}\vspace{-1.5cm}
\centerline{\includegraphics[scale=0.67]{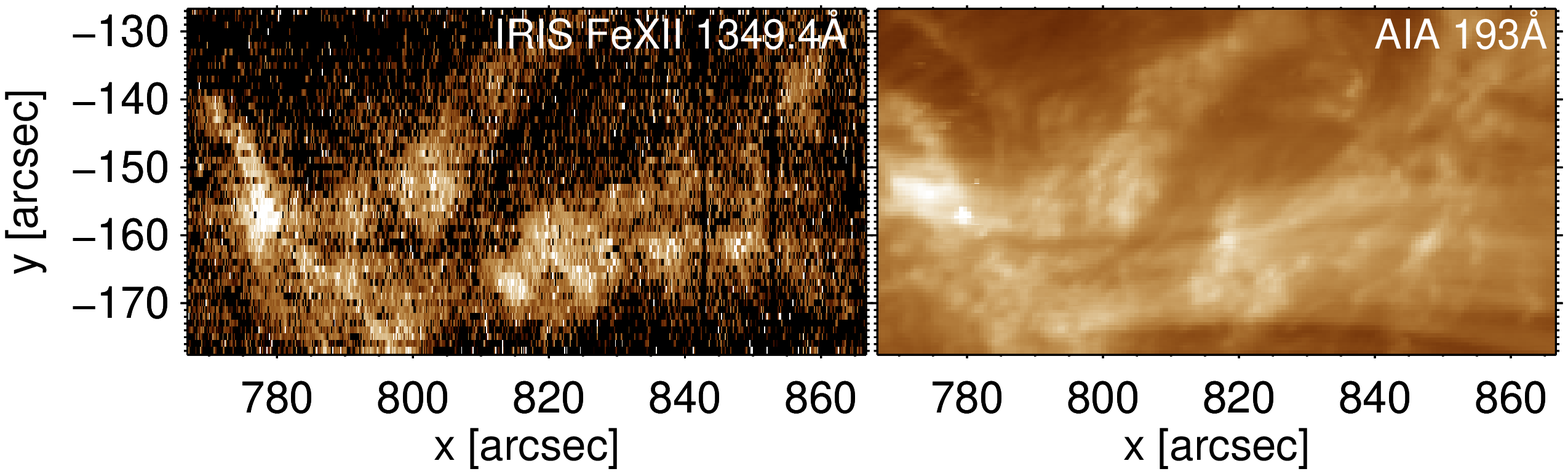}}\vspace{-0.7cm}
\caption{\iris\ and \sdo-\aia\ observations of AR 12014. {\em Top:}
  \iris\ slit-jaw image in the 1400\AA\ passband ({\em left}), and
  \aia\ 193\AA\ image ({\em right}). For a subregion, indicated by the
  box in the \iris\ SJI and \aia\ images, we also show the intensity map in
  the \fexii\ 1349\AA\ line (from Gaussian fit to the \iris\ spectra;
  {\em bottom left}), and the corresponding composite \aia\ 193\AA\
  image (built by selecting, for each horizontal slice of the \iris\
  f.o.v.\ the \aia\ data closest in time; {\em bottom right}).  
  \label{fig:moss}} 
\end{figure*}

For the second dataset, shown in Figure~\ref{fig:moss}, \iris\
observed AR 12014 (2014-03-29 23:27-02:14UT), with 64-step sparse
rasters (i.e., raster steps are $\sim 1$\arcsec), 30s exposure times,
no spatial binning, FUV spectral binning $\times 2$, lossless
compression (OBSID 3810263243), and a roll angle of -90\deg.
Slit-jaw images (SJI) were obtained in the 1330\AA, and 1400\AA\
passbands with a cadence of $\sim 62$s in each passband. In this
dataset the FUV and NUV spectra have $\Delta \lambda \sim 0.025$\AA,
and a f.o.v.\ of $\sim 63$\arcsec$\times 175$\arcsec.    
For this dataset we have coordinated \hinode\ observations, and in
this paper we are especially focused on spectral observations with
\eis.
The \eis\ instrument \citep{Culhane07} observes two wavelength ranges
(171-212\AA\ and 245-291\AA) with a spectral resolution of $\sim
0.023$\AA, and provides solar imaging by stepping (W to E) the slit
(oriented in the N-S direction) over a region of the Sun. 
The \eis\ observations we analyze here (2014-03-30 01:36-02:02) are
characterized by a field of view (f.o.v.) of
100\arcsec$\times$240\arcsec, slit width of 2\arcsec, and exposure
time of 30~s at each step (study acronym PRY\_footpoints\_lite).  Of
the large list of strong lines included in the \eis\ study, in this
paper we will focus on the \fexii\ transitions (with particular
emphasis on the 195.119\AA\ line), and the \fexiii\ transitions
(202.044\AA, 203.772\AA+203.796\AA) which provide useful density
diagnostics \citep{Young09}.

We use \iris\ calibrated level 2 data, which have been processed for
dark current, flat field, and geometrical corrections
\citep{depontieu14}. To correct the absolute wavelength scale in the
FUV, we use the neutral line of \oi\ at 1355.6\AA\ as our zero
velocity reference since it is expected to have, on average, intrinsic
velocity of less than 1 km~s$^{-1}$ (when averaged along the slit;
\citealt{depontieu14}).  
  
For both datasets we analyze simultaneous \aia\ \citep{Lemen12} level
1.5 data, processed for bad-pixel removal, despiking, flat-fielding,
and image registration (coalignment among the different passbands, and
adjustments of roll angle and plate scales), and \hinode-\xrt\
\citep{Golub07} data, processed with the xrt\_prep routine available
in SolarSoft. 
\aia\ observes the full Sun at high temporal cadence ($\sim 12$s) and
spatial resolution (pixel size $\sim 0.6$\arcsec)  in several narrow
EUV passbands sampling the coronal in a wide temperature range
\citep{Lemen12,Boerner12,Boerner14}.  Here we focus on two of the
\aia\ passbands: the 193\AA\ band, in which \fexii\ emission is
generally dominant (see e.g., \citealt{MartinezSykora11}), and the
94\AA\ band which is sensitive to hot plasma ($\log T[$K$] \sim
6.6-7$) because it includes a \fexviii\ line \citep[e.g.,][]{Testa12}
with strong emission in the core of active regions (e.g.,
\citealt{reale11,Testa12b}).  \xrt\ observes the solar corona in soft
X-ray broad bands, sensitive to a wide temperature range ($\log T[$K$]
\sim 6-7.5$; e.g., \citealt{Reale09}).  

The \eis\ data are processed with SolarSoft routines that: (a) remove
the CCD dark current, cosmic-ray strikes on the CCD, (b) take into
account hot, warm, and dusty pixels, (c) apply radiometric calibration
to convert the data to physical units, and (d) apply wavelength
corrections for orbital variations and slit tilt. 

We coaligned the datasets by applying a standard cross-correlation
routine (tr\_get\_disp.pro, which is part of the IDL SolarSoftware
package), to the \fexii\ emission maps obtained with the different
instruments. For \aia\ we build 193\AA\ synthetic rasters (composite
image, hereafter) by using for each stripe along a raster position
(horizontal stripe for the \iris\ observations we selected, and
vertical for \eis) the corresponding \aia\ data closest in time to 
when the spectrograph slit was at that location. Given the
uncertainties in the absolute pointing of the instruments, a few
iterations are needed to find a good coalignment.
In Figures~\ref{fig:loops} and \ref{fig:moss} we show for each dataset
one slit-jaw image and the \aia\ 193\AA\ image closest in time, and,
for a subregion scanned by the \iris\ slit, we show the raster
intensity map in the \iris\ \fexii\ 1349.4\AA\ line, and the
corresponding \aia\ 193\AA\ composite image.

\begin{figure*}[!ht]
\vspace{0.7cm}
\centerline{\includegraphics[scale=0.5]{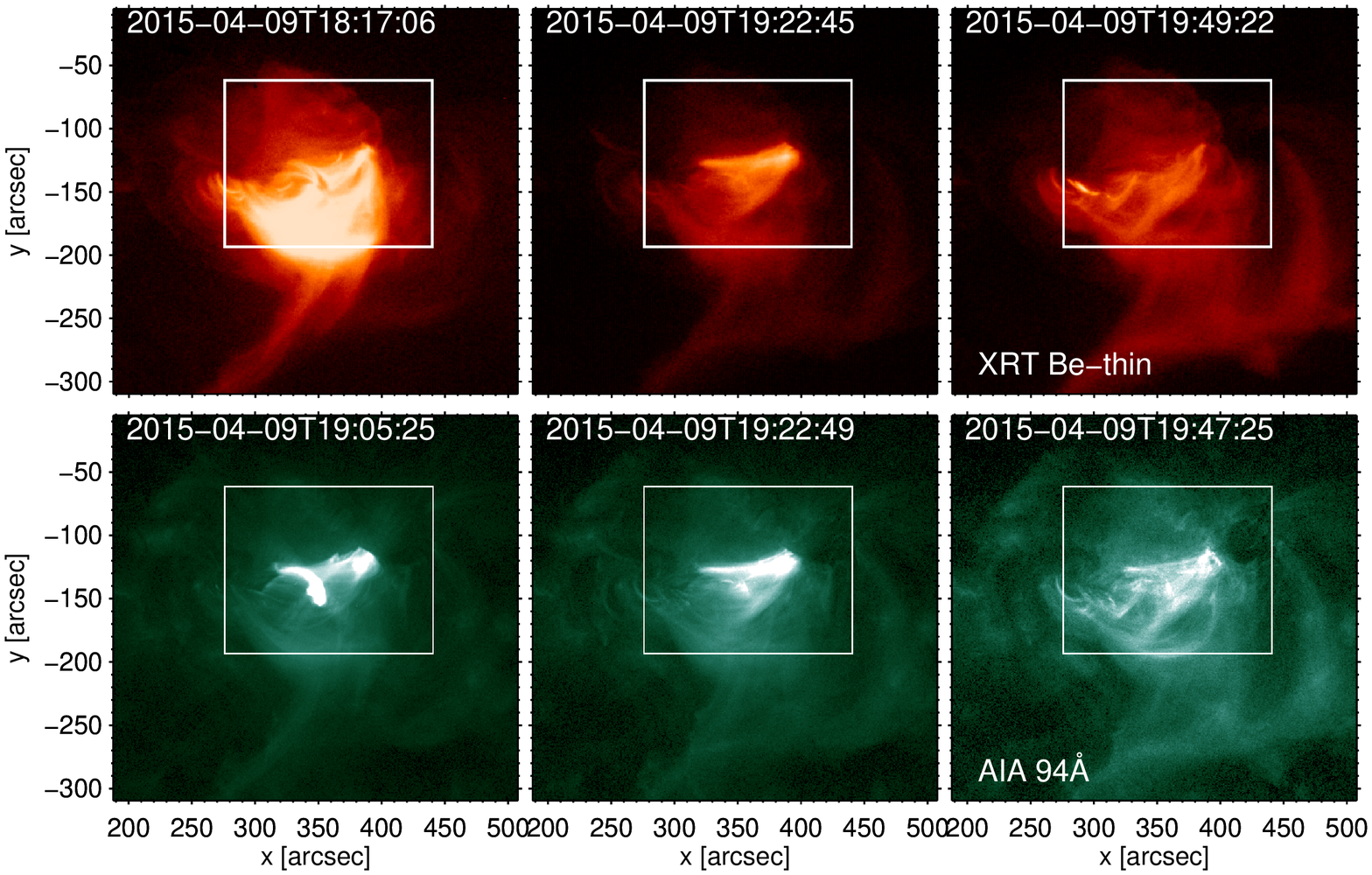}
  \includegraphics[scale=0.5]{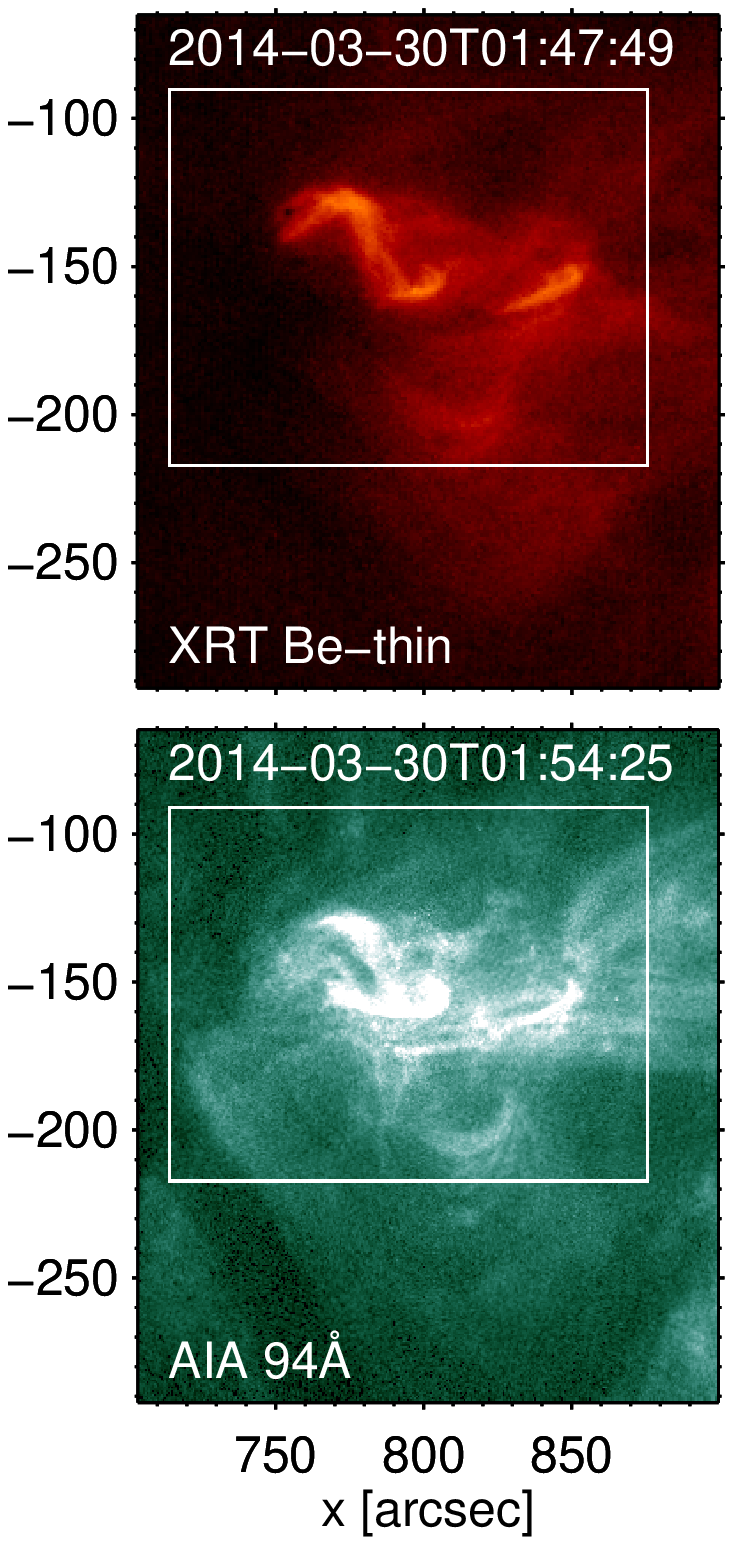}}%\vspace{-0.2cm}
\caption{\hinode-\xrt\ (Be-thin passband) and \aia\ 94\AA\ images
  showing the hot plasma present in the active regions we study
  (Figures~\ref{fig:loops} and \ref{fig:moss}). For the dataset of
  2015-04-09 we show the hot plasma at three different times: the peak
  of the X-ray/94\AA\ emission (first column), the peak of the hot
  emission in the loops that we see at a later time in \fexii\ (second
  column), and a time closer to the \iris\ and \aia\ images shown in
  Figure~\ref{fig:loops} (third column). 
  For the moss dataset of 2014-03-30 we show the hot emission at a
  time close to the \iris\ and \aia\ images of Figure~\ref{fig:moss}
  (fourth column).
  The white boxes show the f.o.v.\ of the \iris\ SJI and \aia\ 193\AA\
  images shown in the top panels of Figure~\ref{fig:loops} and
  \ref{fig:moss}.
  \label{fig:hot}} 
\end{figure*}

For the 2015-04-09 dataset the \iris\ \fexii\ intensity map shows the
presence of elongated coronal structures and additional \fexii\ fuzzy
emission both above and below the bright loops. 
The \aia\ 193\AA\ observations (Fig.~\ref{fig:loops}, and
movies) show that the \fexii\ emission visible with \iris\ corresponds
to bright transient loops in the core of the active region and some
bright moss.  
Inspection of additional \aia\ and \hinode-\xrt\ data shows that these
bright loops appear in the late phases of the evolution of a C6.2
flare (GOES peak at 18:51UT; see Figure~\ref{fig:hot}). 

In the 2014-03-29 dataset, the \fexii\ emission, in both \iris\ and
\aia, is generally dominated by emission of bright moss, due to the
high density of transition region plasma
\citep[e.g.,][]{Fletcher99}. The \aia\ 94\AA\ images and the
\hinode-\xrt\ images confirm the presence of hot dynamic loops in this
active region (Figure~\ref{fig:hot}).  

\section{Analysis Methods and Results}
\label{s:results}

The \iris\ \fexii\ spectral line properties are obtained by fitting
the spectra with a single Gaussian (plus a constant background). 
The intensity maps obtained for the two datasets are shown in
Figure~\ref{fig:loops} and \ref{fig:moss}, and generally show good
correspondence with the brightest regions of the corresponding \aia\
193\AA\ composite images.  

From the spectral fit, we derive the non-thermal line width 
$w_{\rm  nth} = \sqrt{w^{2}_{1/e} - w^{2}_{th} - w^{2}_{instr}}$,
where $w_{\rm 1/e}$ is the measured $1/e$ spectral line width (i.e.,
$\sqrt{2} \times$ the Gaussian $\sigma$), $w_{\rm th}$ is the thermal
line width, and $w_{\rm  instr}$ is the instrumental line width.
The thermal line width is $\sqrt{2 K_{\rm B} T/m_{\rm ion}}$, and for
the temperature of peak formation of \fexii\ ($\log T[K] = 6.2$) it
corresponds to $\sim 21.7$ km~s$^{-1}$. 
The instrumental line width of \iris\ is of the order of 4~km~s$^{-1}$
\citep{depontieu14}, while for our \eis\ dataset the instrumental line
width is of order of $60-65$~km~s$^{-1}$ and variable along the slit (we used
the eis\_slit\_width SolarSoft routine to calculate the instrumental width).

\begin{figure*}[!ht]
\centerline{\includegraphics[scale=0.6]{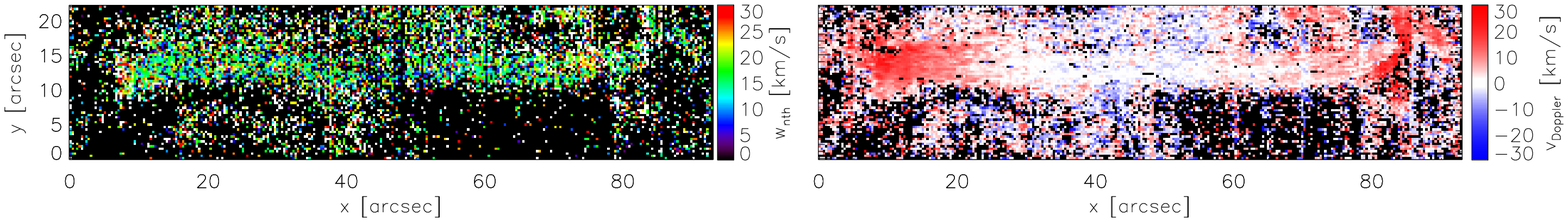}}\vspace{-0.3cm}
\centerline{\hspace{0.1cm}\includegraphics[scale=0.6]{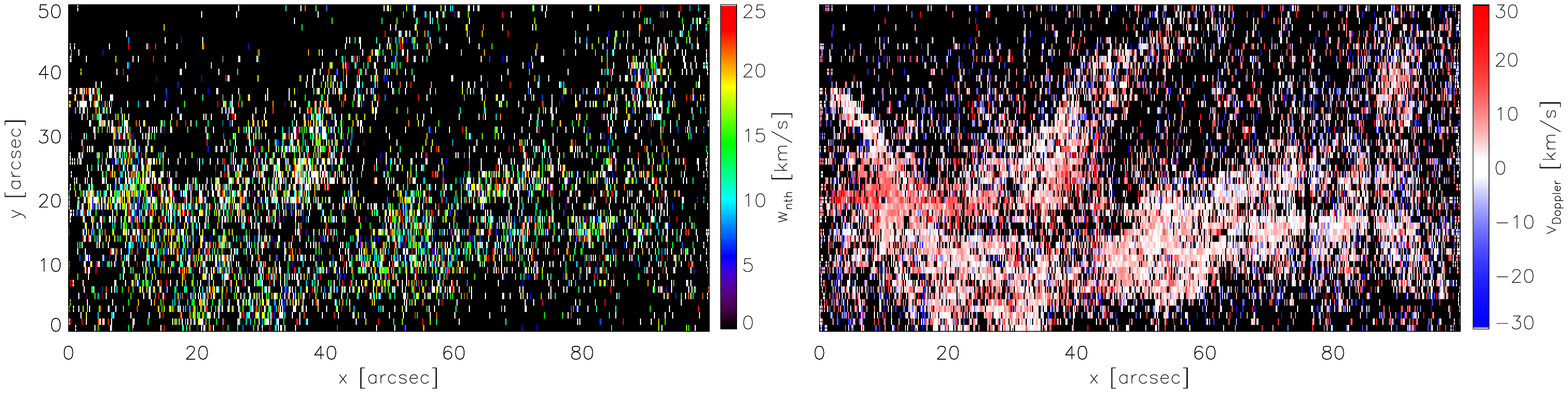}}
\caption{Maps of \fexii\ non-thermal line width ({\em left column}) and Doppler
  velocity ({\em right column}) from Gaussian fit to the 1349\AA\ line
  in the \iris\ spectra, for the two datasets of Fig.~\ref{fig:loops}
  ({\em top row}) and \ref{fig:moss} ({\em bottom row}).
  In the Doppler velocity plots blue and red correspond to flows
  towards and away from the observer, respectively.
  In all plots black is used for pixels where the low S/N
  prevents a reliable determination of the velocity.
  In the non-thermal velocity plots white is used for pixels in which
  the \wnth\ is larger than the upper limit of the velocity scale
  (i.e., 30 and 25~km~s$^{-1}$, for the upper and lower panel
  respectively).   
  \label{fig:fe12_iris}} 
\end{figure*}

\begin{figure}[!ht] 
\centerline{\includegraphics[scale=0.55]{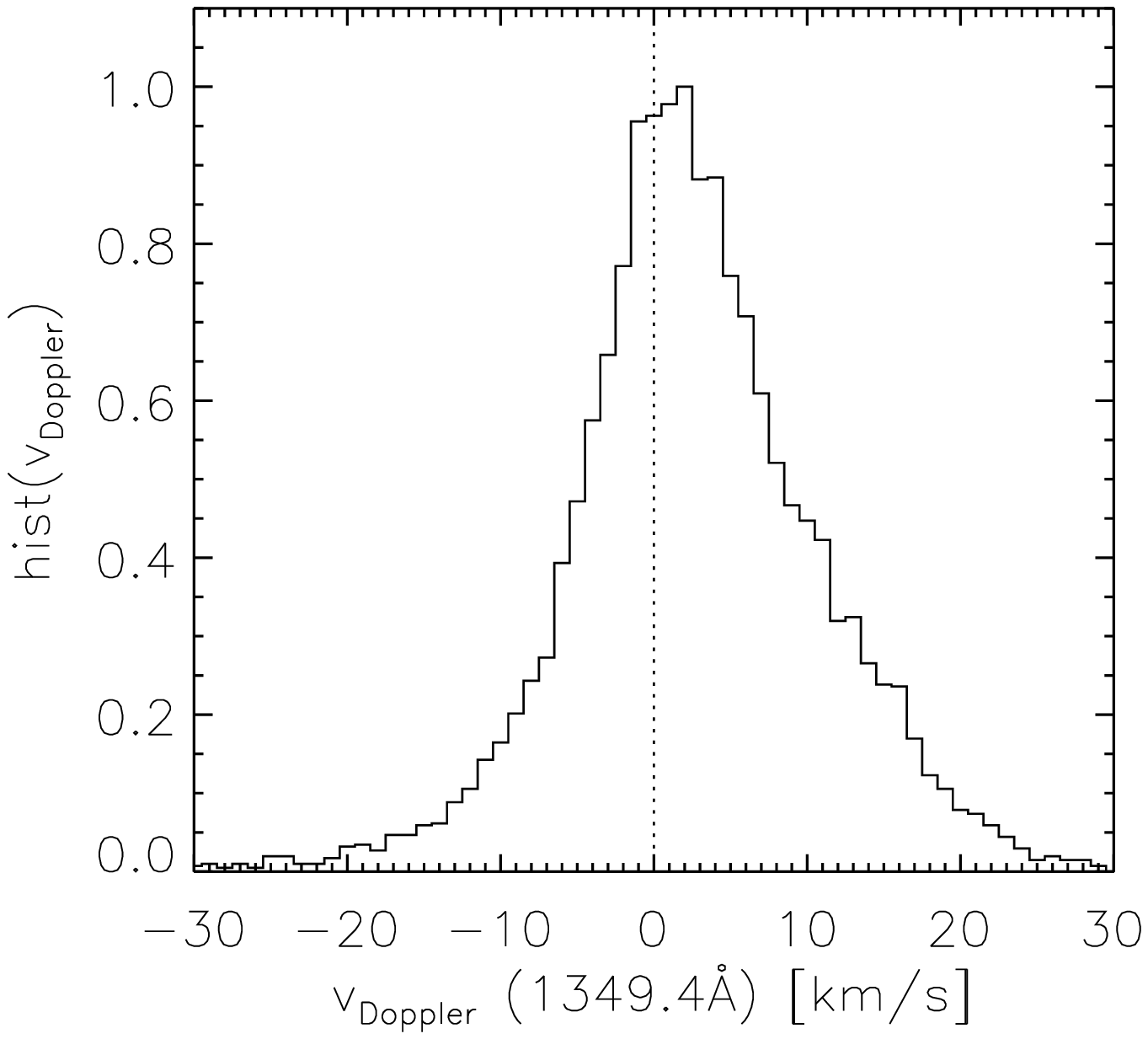}}\vspace{-0.5cm}
\centerline{\includegraphics[scale=0.55]{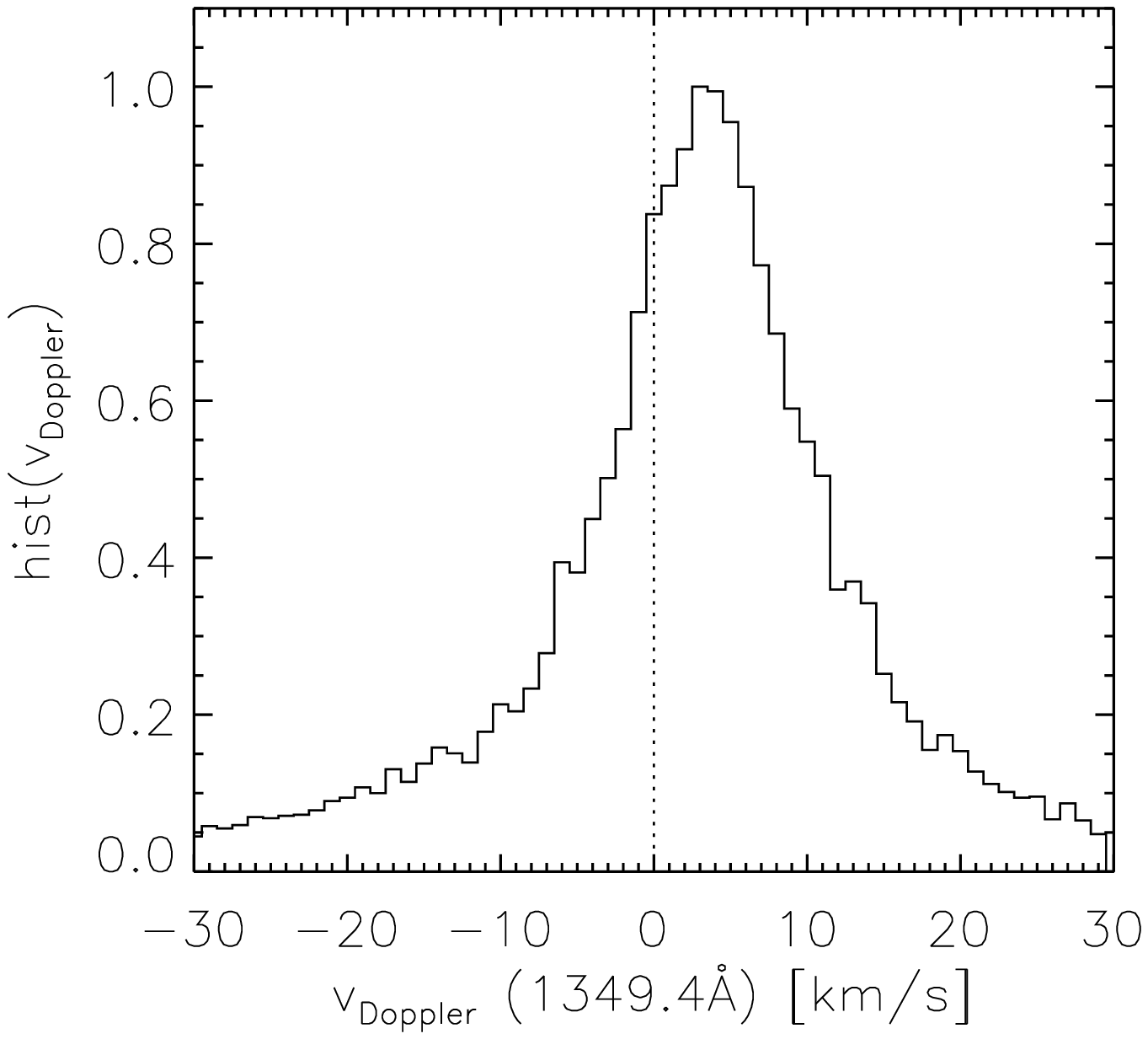}}
\caption{Normalized histogram of Doppler shifts in the 1349\AA\
  \iris\ \fexii\ line (right panels of Figure~\ref{fig:fe12_iris}),
  for the 2015-04-09 observations including the post-flare loops ({\em
    top}), and the moss observation of 2014-03-30 ({\em bottom}). 
  The dotted line indicates zero velocity,
  positive velocities correspond to redshifts (i.e., velocity toward
  the solar photosphere), and negative velocities correspond to
  blueshifts (i.e., outward velocities, toward the observer). 
  The absolute velocities are calibrated using the \oi\ line at
    1355.6\AA\ (see sec.\ref{s:data}).
  \label{fig:vdoppler}} 
\end{figure}

The non-thermal line width and Doppler shift for the two selected
\iris\ datasets are shown in Figure~\ref{fig:fe12_iris}. The
non-thermal line width maps show that the typical residual observed
non-thermal velocities are modest, typically of the order of
$15-20$~km~s$^{-1}$.  The Doppler shift map of the 2015 dataset shows
for the postflare loops significant redshift, more pronounced on the
eastern footpoints, suggesting draining of plasma in the late phases
of the flare. Moss is also visible in that observation and it shows more
a mix of blue and redshifts. 
The histogram of Doppler velocities (top panel of
  Figure~\ref{fig:vdoppler}) is asymmetric, with respect to the peak,
  and shows a significant red shoulder due to the post-flare loop footpoints.
In the observations of 2014 the moss \fexii\ emission is also
characterized by a mix of blue and redshifts, and the distribution
(bottom panel of Figure~\ref{fig:vdoppler}) is peaked around
$\sim +3$~km~s$^{-1}$ (i.e., slightly redshifted), and it is
significantly more symmetric than the 2015 dataset which includes the
post-flare loops.  
We note that the line shifts we observe reveal the line of sight
component of the plasma velocity: since for the moss observations of
2014-03-29 the active region is far from disk center
(770\arcsec,-150\arcsec) the actual velocity of the \fexii\ emitting
plasma could be larger than our measured values of few km~s$^{-1}$. 

\begin{figure*}[!ht]
\centerline{\includegraphics[scale=0.7]{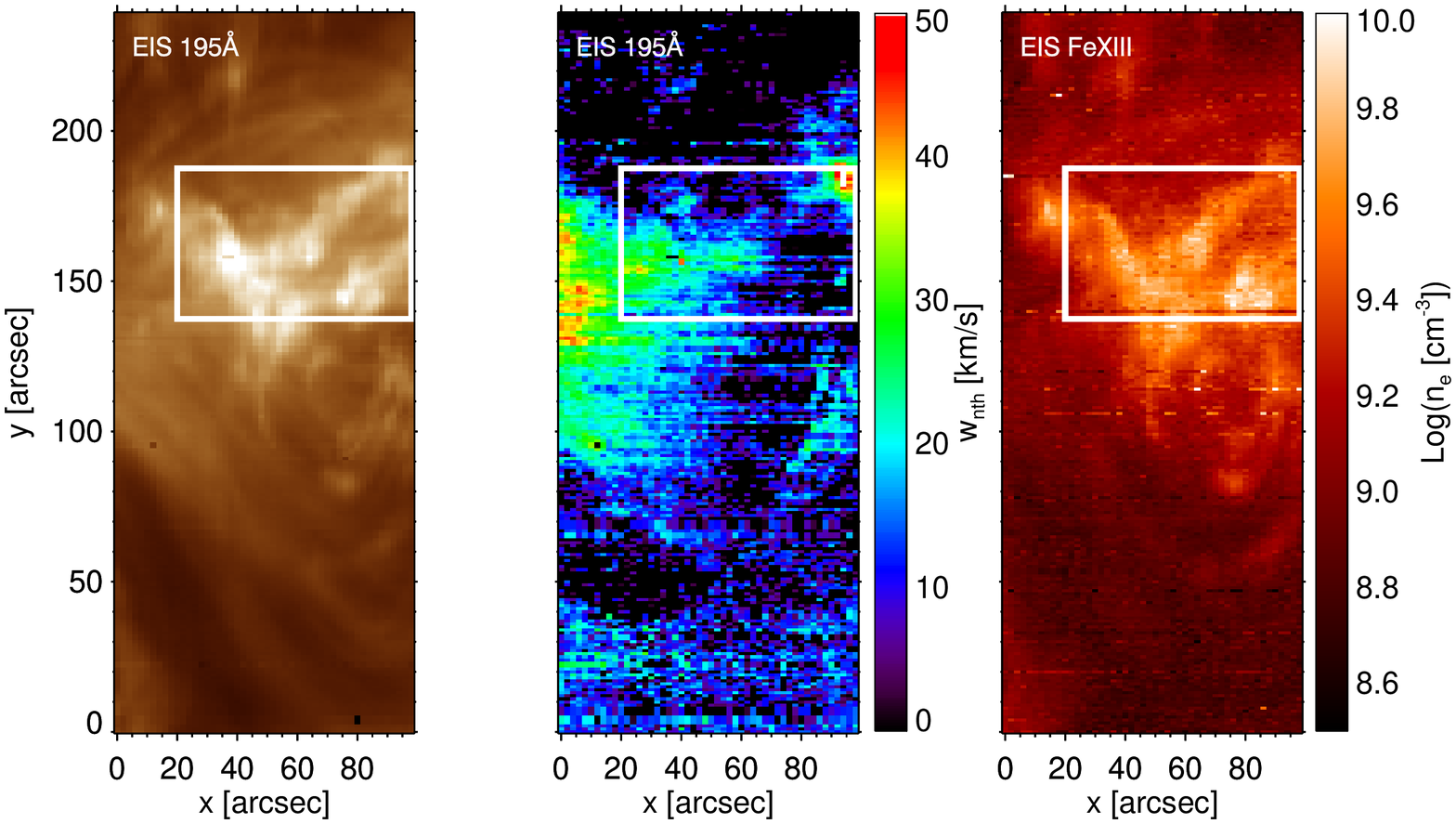}}
\caption{\hinode-\eis\ observations of AR 12014 simultaneous to the
  \iris\ and \aia\ observations shown in Fig.~\ref{fig:moss}.
  We show the maps of the \eis\ \fexii\ 195.12\AA\ line intensities ({\em
    left}) and non-thermal velocities ({\em middle}), and the plasma
  electron densities obtained from the measured ratios of the 202\AA\
  and 203\AA\ \fexiii\ \eis\ lines ({\em rigth}).   The box indicates
  the region of overlap between the \eis\ and the \iris\ spectral
  observations. 
  \label{fig:fe12_eis}} 
\end{figure*}

We derive the \eis\ \fexii\ 195.119\AA\ intensity and non-thermal width
by fitting the \eis\ spectral data with a double Gaussian to remove
the blend with the weaker \fexii\ line at 195.179\AA. We impose that
the value of the $\sigma$ of the two Gaussian components is the same,
and we fix the wavelength separation $\Delta \lambda$ of the two line
centroids to the separation of the theoretical wavelengths as derived
from CHIANTI (see also \citealt{Young09}).   
The maps of line intensity and non-thermal width are shown in
Figure~\ref{fig:fe12_eis}.  In the region of overlap of the \eis\ and
\iris\ f.o.v.\ the \fexii\ 195\AA\ intensity morphology is very
similar to the \iris\ observed 1349\AA\ emission, even though the
\iris\ and \eis\ observations are not strictly simultaneous at all
locations. \eis\ rasters W to E, with the slit in the N-S
direction. For this observation \iris\ is rolled by -90 degrees,
i.e., 90\deg\ counterclockwise with respect to the zero degree roll (roll=0
corresponds to a N-S orientation of the slit). In this observing
sequence \iris\ rasters S to N (when roll=0 \iris\ rasters E to W).  
Moss emission is typically observed to be relatively steady over timescales
long compared to the $\sim 30$~min of the \iris\ and \eis\ rasters we
analyze here (\citealt{Antiochos03,Brooks09,Tripathi10}, though see
also \citealt{testa13,Testa14}), therefore we expect the
non-simultaneity of the spectra at each location to have limited
effects on the \fexii\ spectral properties derived with \iris\ and
\eis.  We discuss more, later in this section, the extent and effects
of the non-simultaneity of the observations with the two instruments.

The largest non-thermal velocities of the \eis\ 195\AA\ line are
observed at the footpoints of large fan loops while moss is
characterized by significant smaller non-thermal motions.   
This is in agreement with previous studies \citep[e.g.,][]{doschek08}.
In Figure~\ref{fig:fe12_eis} we also show a map of the plasma electron
density derived from the ratio of the 202\AA\ and 203\AA\ \fexiii\
lines \citep[e.g.,][]{Young09}. This shows the high density of plasma
in moss regions \citep[e.g.,][]{Fletcher99}.

\begin{figure*}[!ht]
\centerline{\includegraphics[scale=0.7]{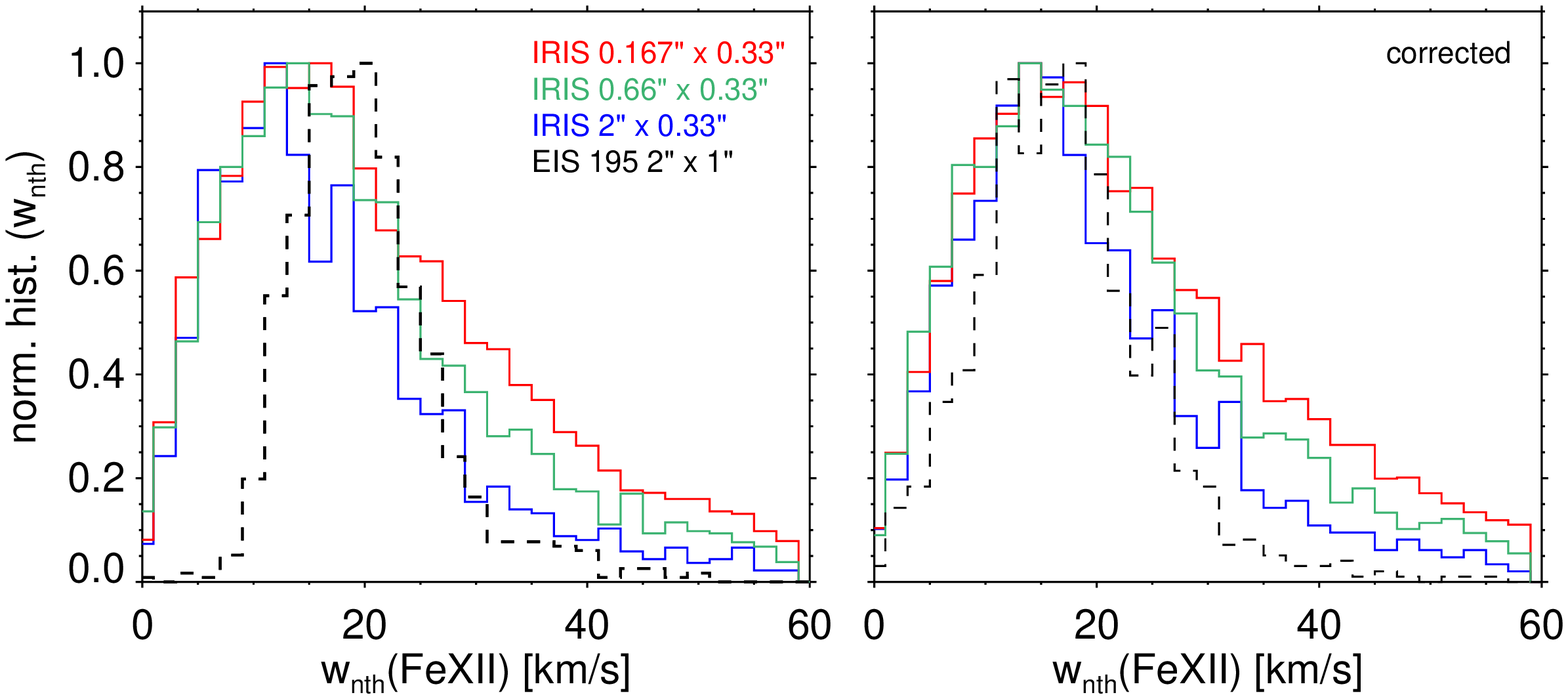}}
\caption{Normalized histograms of non-thermal line width for the moss
  region where both \iris\ and \eis\ data are available. For \iris\ we
  show the histograms for three different rebin levels (in the slit
  direction, i.e., E-W): the original spatial resolution
  ($0.167$\arcsec$\times 0.33$\arcsec; red line), and a rebin of a
  factor 4 ($0.66$\arcsec$\times 0.33$\arcsec; green) and 12
  ($2$\arcsec$\times 0.33$\arcsec; blue).  The dashed lines
    corresponds to the \eis\ data.
  The right panel shows the histograms obtained by correcting for
  instrumental effects of both \iris\ and \eis\ leading to systematic
  error in the measurements of line width (see text and
  Figure~\ref{fig:fe12_corr}).  
  \label{fig:fe12_hist}} 
\end{figure*}

In Figure~\ref{fig:fe12_hist} ({\em left panel}) we compare the
distributions of non-thermal velocities in the \fexii\ emission
observed with \iris\ and with \eis. For \iris\ we computed the
distributions for different spatial resolution: (a) for the
\iris\ original spatial resolution ($0.167$\arcsec$\times
0.33$\arcsec; the width of the \iris\ slit is
$0.33$\arcsec), (b) for a rebin by a factor 4 along the slit (i.e.,
$0.66$\arcsec$\times 0.33$\arcsec\ macropixels), and (c) for a rebin
by a factor 12 ($2$\arcsec$\times 0.33$\arcsec\ macropixels, i.e.,
only 3 times smaller than the \eis\ pixels). 
We do not integrate further because the \iris\ raster for this OBSID
is not ``dense'', i.e., there are $\sim$1\arcsec\ gaps in between raster
positions, and even if we did sum consecutive raster positions they
would not be simultaneous. 
The distributions of Figure~\ref{fig:fe12_hist} show that for larger
rebinning factors the \iris\ \wnth\ distributions have a smaller tail
at large values. This can be expected if there is significant
sub-arcsec structuring and the structures with large \wnth\ are
generally not concentrated close to each other, and/or in the vicinity
of locations with brighter narrow lines.  
The \eis\ distribution is markedly narrower than the \iris\
distributions, and it also peaks at larger \wnth\ values ($\sim
20$~km~s$^{-1}$, vs.\ $10-15$~km~s$^{-1}$ for \iris). 

\begin{figure*}[!ht]
\centerline{\includegraphics[scale=0.7]{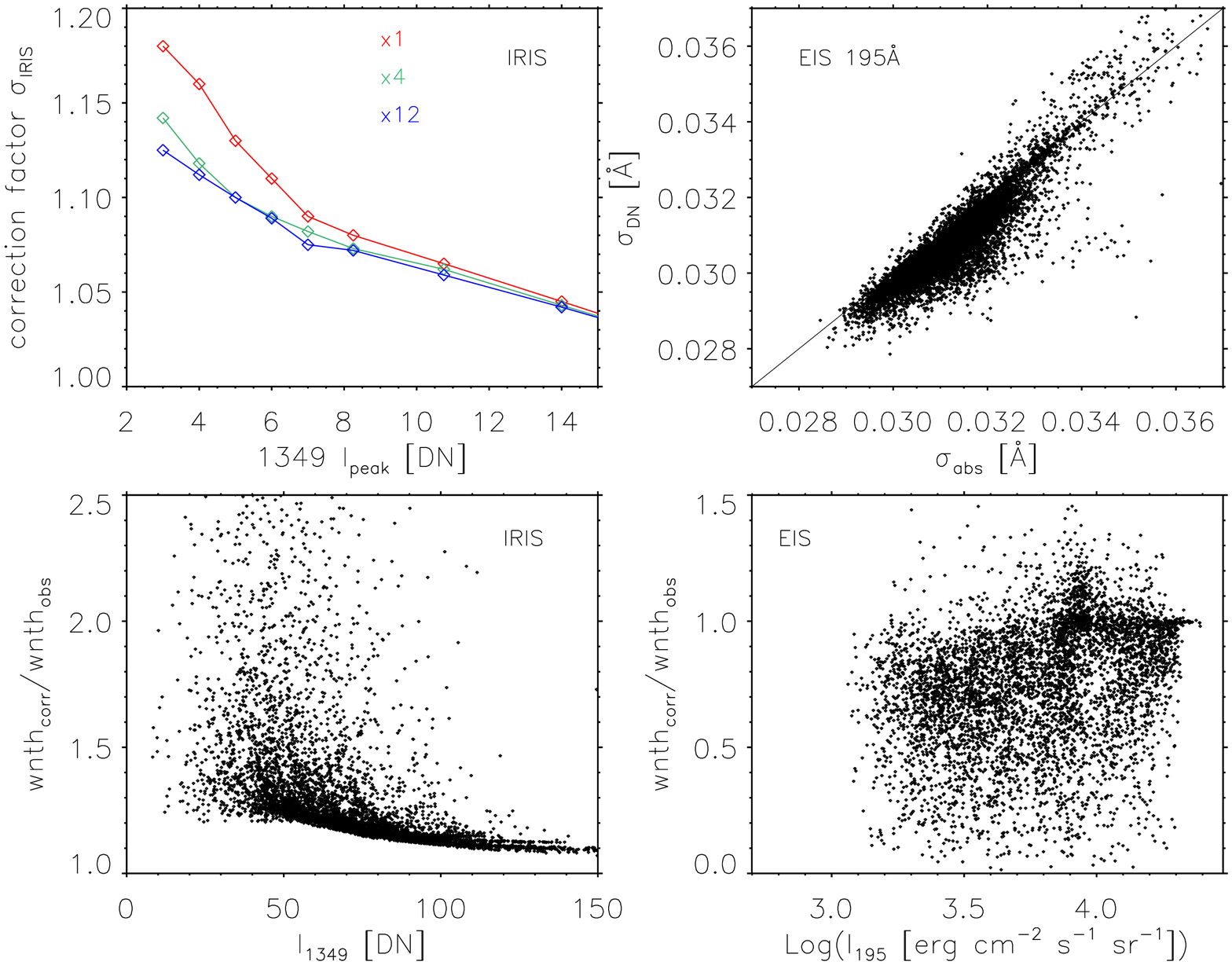}}
\caption{Corrections of instrumental effects leading to systematic
  error in line width measurements. {\em Top left:} Correction factor for
  the Gaussian $\sigma$ in the fit to the \iris\ \fexii\ spectral
  line, as obtained from simulations (see text). {\em Top right:}
  Gaussian $\sigma$ from the fit of \eis\ \fexii\ 195\AA\ for spectra
  in DN units (y axis) and absolute units (x axis). {\em Bottom:}
  ratio of the corrected to uncorrected non-thermal widths as a
  function of line intensity for the \iris\ 1349\AA\ line ({\em
    left}), and the \eis\ 195\AA\ line ({\em right}), for the region
  of overlap of the f.o.v.\ of \iris\ and \eis\ spectral
  observations (see Fig.~\ref{fig:fe12_eis}). 
  \label{fig:fe12_corr}} 
\end{figure*}

The histograms of \wnth\ from \eis\ and from \iris\ for the largest rebinning
factor present significant differences, whereas one would expect them
to be similar. We therefore investigated possible causes for the
observed large difference.  
For \iris\ we investigated possible effects of the typically low
signal-to-noise ratio on the determination of \wnth: in pixels with
low \fexii\ signal, the wing of the Gaussian spectral line can have
signal that is lower than the digitization threshold (i.e., if the signal in a
spectral bin is below 1~data number [DN], the bin gets assigned a
value of 0~DN), possibly leading to a systematic underestimate of the
Gaussian $\sigma$. 
To test this hypothesis we run Monte Carlo simulations of \iris\
spectra, varying the following parameters: (1) peak intensity (from 3
to 15 DN; pixels with smaller peak intensities are anyway filtered out
because the fits are not reliable enough), (2) \wnth\ (from 5 to
20~km~s$^{-1}$), (3) background value (1-16~DN/pix). 
For each set of parameters we run 3000 simulations applying Poisson
noise to each spectral pixel, and derive the correction factor
(averaged over the 3000 simulations) which multiplied by the measured
$\sigma$ gives the ``true'' Gaussian $\sigma$ of the input model. 
We find that low SNR does indeed cause a slight systematic
underestimation (typically by 5-15\%) of $\sigma$ (and therefore of \wnth), and
the peak intensity is the parameter the effect mainly depends on. In
Figure~\ref{fig:fe12_corr} ({\em top left panel}) we show the average
correction factor for the Gaussian $\sigma$, by which the measured
$\sigma$ should be multiplied to determine the actual width in the
\iris\ spectral fit, as a function of the peak intensity, and for
different rebinning factors. In Figure~\ref{fig:fe12_corr} we also
show the resulting effect on the \wnth\ by plotting the ratio of the
corrected to ``uncorrected'' \wnth\ as a function of measured 1349\AA\ line
intensity ({\em bottom left panel}).  
This figure shows that the modest corrections in the Gaussian $\sigma$
can propagate to significant corrections for the non-thermal
velocities. 

 For \eis\ we explored the effect of the absolute calibration (D.\
Brooks, priv.\ comm., and \citealt{brooks15}) by comparing
the Gaussian $\sigma$ obtained through the fit to the \eis\ spectra in
physical units (erg~cm$^{-2}$~s$^{-1}$~sr$^{-1}$) to the one obtained
by fitting \eis\ spectra in DN units (obtained by applying the
eis\_prep SolarSoft routine with the /noabs keyword). 
The comparison ({\em top right panel} of Figure~\ref{fig:fe12_corr})
shows that the absolute calibration leads to a systematic overestimate
of the line width for most pixels. While the effect in the measured
$\sigma$ is relatively small (typically $\lesssim 10$\%), this
propagates to significant changes of \wnth. We show in the bottom
right panel of Figure~\ref{fig:fe12_corr} the ratio of the corrected
to ``uncorrected'' \wnth\ as a function of the 195\AA\ line
intensity. 

The \iris\ and \eis\ histograms of \wnth, corrected for the above
discussed instrumental effects, are shown in the right panel of
Figure~\ref{fig:fe12_hist}. As explained above, the correction of
\iris\ \wnth\ leads to histograms with significantly smaller numbers
of pixels with low \wnth, and the peaks of the histograms are around
$15$~km~s$^{-1}$.  The correction of the \eis\ \wnth\ shifts the
histogram to smaller \wnth\ values, and makes the distribution 
similar, both in terms of peak and width, to the (corrected) \iris\
histogram with the largest rebinning factor, therefore largely
reconciling the \iris\ and \eis\ measurements.
This analysis shows the importance of taking into account these
instrumental effects, for a correct interpretation of the spectral
measurements and their comparison among different instruments.
The \wnth\ distributions in Figure~\ref{fig:fe12_hist} show the
effects of spatial resolution: while all distributions at different
spatial resolution are peaked around very similar values, at higher
spatial resolution the distributions are broader both on the low and
on the high side of the range of \wnth\ values. Similar results were
recently found by \cite{depontieu15} for \siiv\ transition region
emission as observed by \iris\ in a variety of solar features (active
regions, quiet sun, coronal holes; see their
Figure~2). \cite{depontieu15} focus on the relative lack of
sensitivity of the peak of the \wnth\ distributions on spatial
resolution, and their possible interpretation includes: broadening
processes occurring along the line-of-sight and/or on spatial scales
smaller than the \iris\ resolution.
The broadening of the \wnth\ distributions at higher spatial
resolution however strongly suggests that the unprecedented \iris\
resolution allows us to resolve at least some structuring of the
non-thermal motions at subarcsecond resolution.

\begin{figure}[!ht]
\centerline{\includegraphics[scale=0.55]{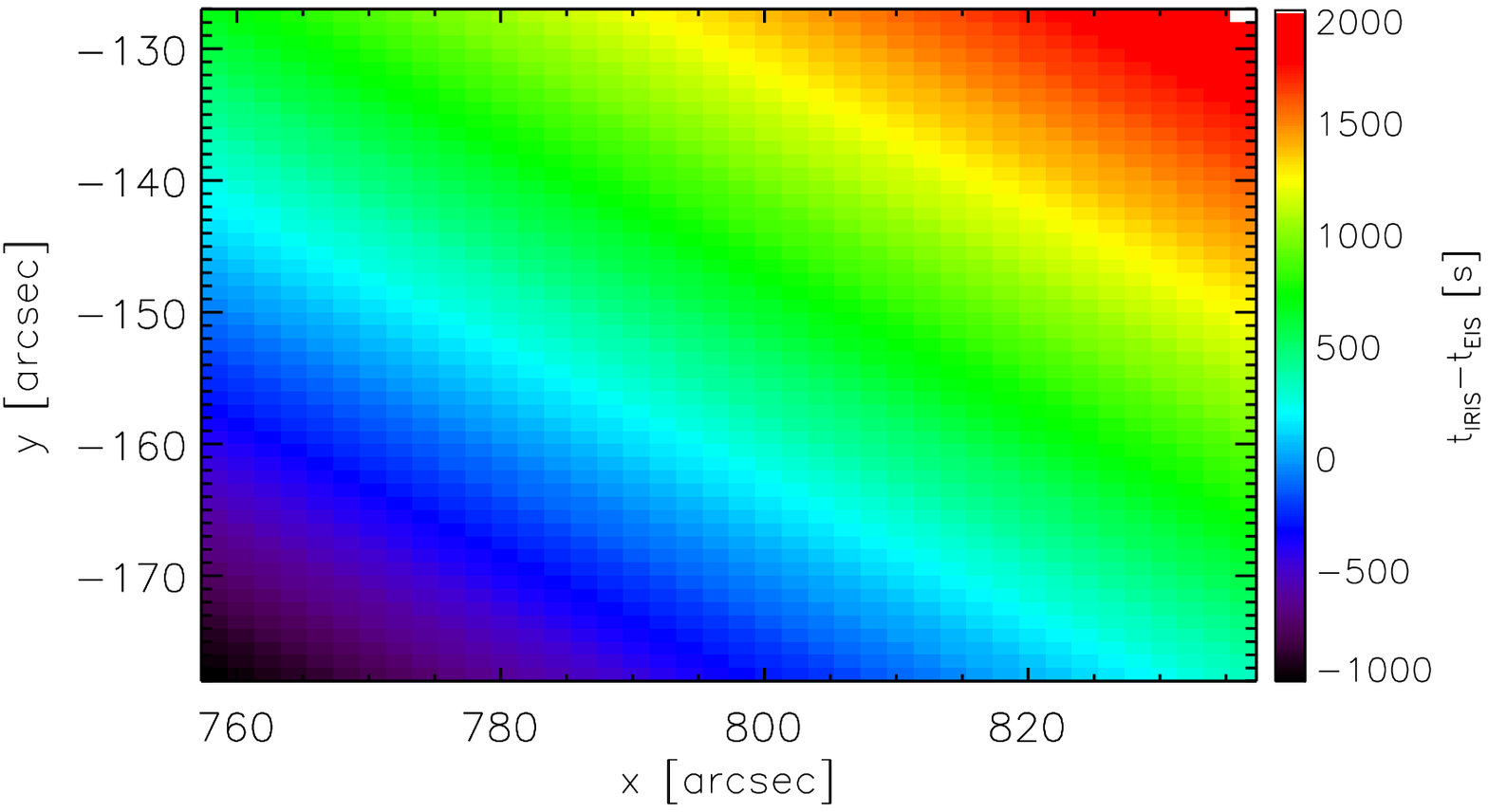}}
\centerline{\includegraphics[scale=0.5]{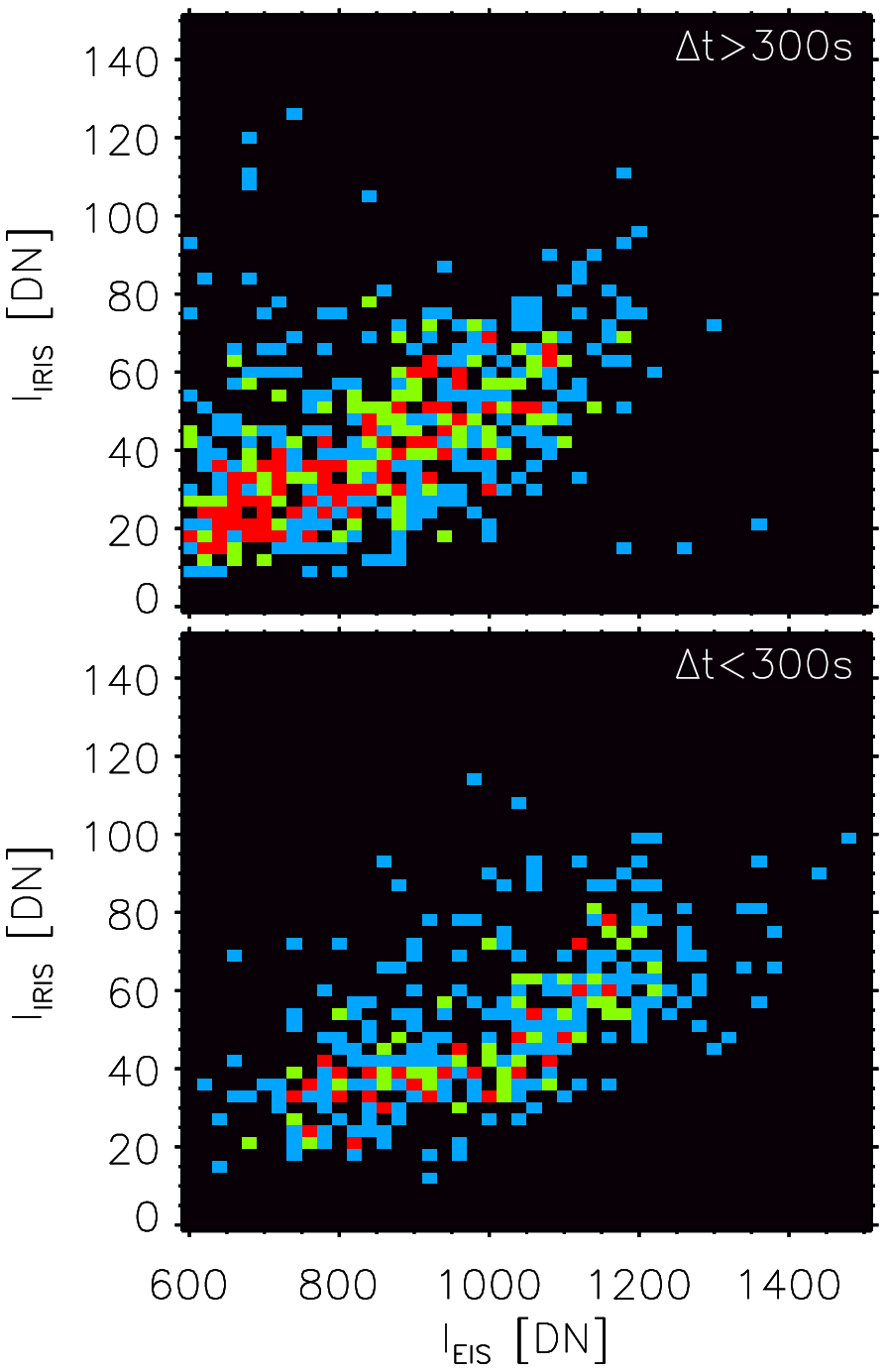}\hspace{-0.5cm}
  \includegraphics[scale=0.5]{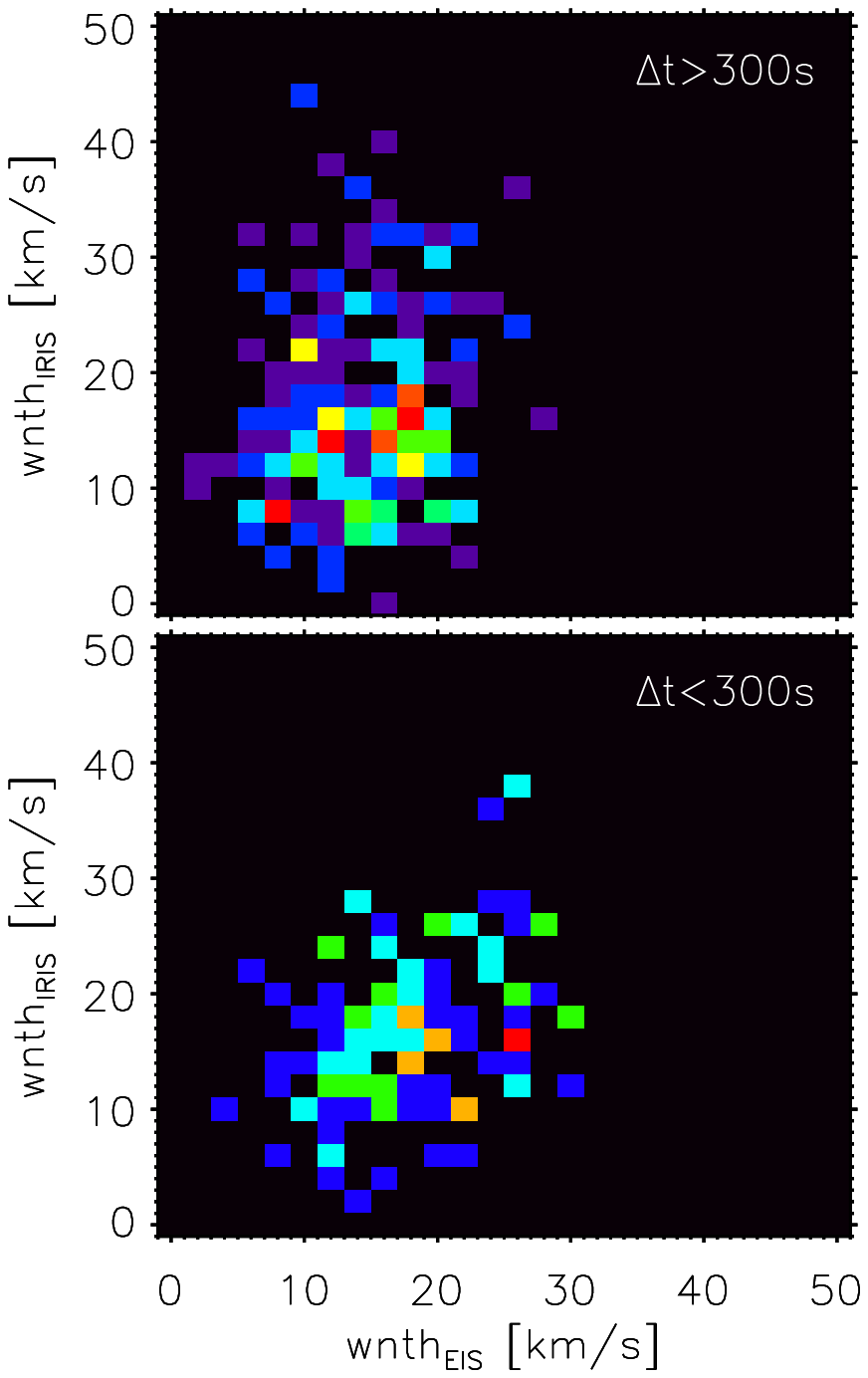}}
\caption{{\em Top panel:} Spatial map of difference in time (in
  seconds) between \iris\ and \eis\ observations in the area of
  overlap of the f.o.v.\ of the two instruments. 
  {\em Middle and bottom panels:} 2-dimensional histograms showing the
  correlation between intensity in \fexii\ lines ({\em left panels})
  and non-thermal line widths ({\em right panels}) as measured by
  \iris\ (1349\AA\ line; spatially integrated by a factor 12, i.e.,
  with a spatial bin of 2\arcsec$\times$0.33\arcsec), and \eis\
  (195\AA\ line, at the original spatial resolution, i.e., with a
  pixel of 2\arcsec$\times$1\arcsec). The middle and bottom panels
  show the 2D histograms (i.e., the two dimensional joint
    probability density function) for the subset of spatial pixels
  where the difference in time between \iris\ and \eis\ observations
  is $\gtrsim 300$~s and $\lesssim 300$~s respectively. We use
    the rainbow color bar (as in top panel), so red corresponds to the
    highest pixel density (followed by orange, yellow, green, cyan,
    blue, black).
\label{fig:correl_eis_iris}} 
\end{figure}

We further compare the \iris\ and \eis\ observed \fexii\ spectral
properties, also taking into consideration the non-simultaneity of the
observations of the two instruments at all locations, due to the
different roll angle and scanning direction of the two instruments.
In Figure~\ref{fig:correl_eis_iris} ({\em upper panel}) we show the
time difference between \iris\ and \eis\ observations, in a spatial
area of overlap of the two fields of view. In the other panels of
Fig.~\ref{fig:correl_eis_iris} we use two-dimensional histograms to
show the correlation of the \iris\ and \eis\ \fexii\ intensity ({\em
  middle and bottom left}) and corrected non-thermal line width ({\em middle and
  bottom right}), for pixels where the time difference is more than
({\em middle}) and less than ({\em bottom}) 300s. 
For \iris\ we use the results obtained on data with a spatial bin of
2\arcsec$\times$0.33\arcsec, to have spatial resolution close to \eis\
resolution. 
These plots show that (a) the \fexii\ intensities show stronger
correlations than the  non-thermal velocities, and that (b) the
correlations are significantly tighter in regions where \iris\ and
\eis\ observations are less than 5~min apart, especially for the
non-thermal velocity. 
The observed differences between \iris\ and \eis\ results are likely
due mostly to the difference in spatial pixels we compare: the \iris\
macropixel we use here (2\arcsec$\times$0.33\arcsec) is three times
smaller in the solar N-S direction than the corresponding \eis\ pixel
Another possible contributing factor to
observed difference between \iris\ and \eis\ is that the \eis\ \fexii\
195\AA\ line suffers significant absorption from cool chromospheric
material \citep[e.g.,][]{depontieu09} which does not affect the \iris\
1349\AA\ line.

\begin{figure*}[!ht]
\centerline{\includegraphics[scale=0.45]{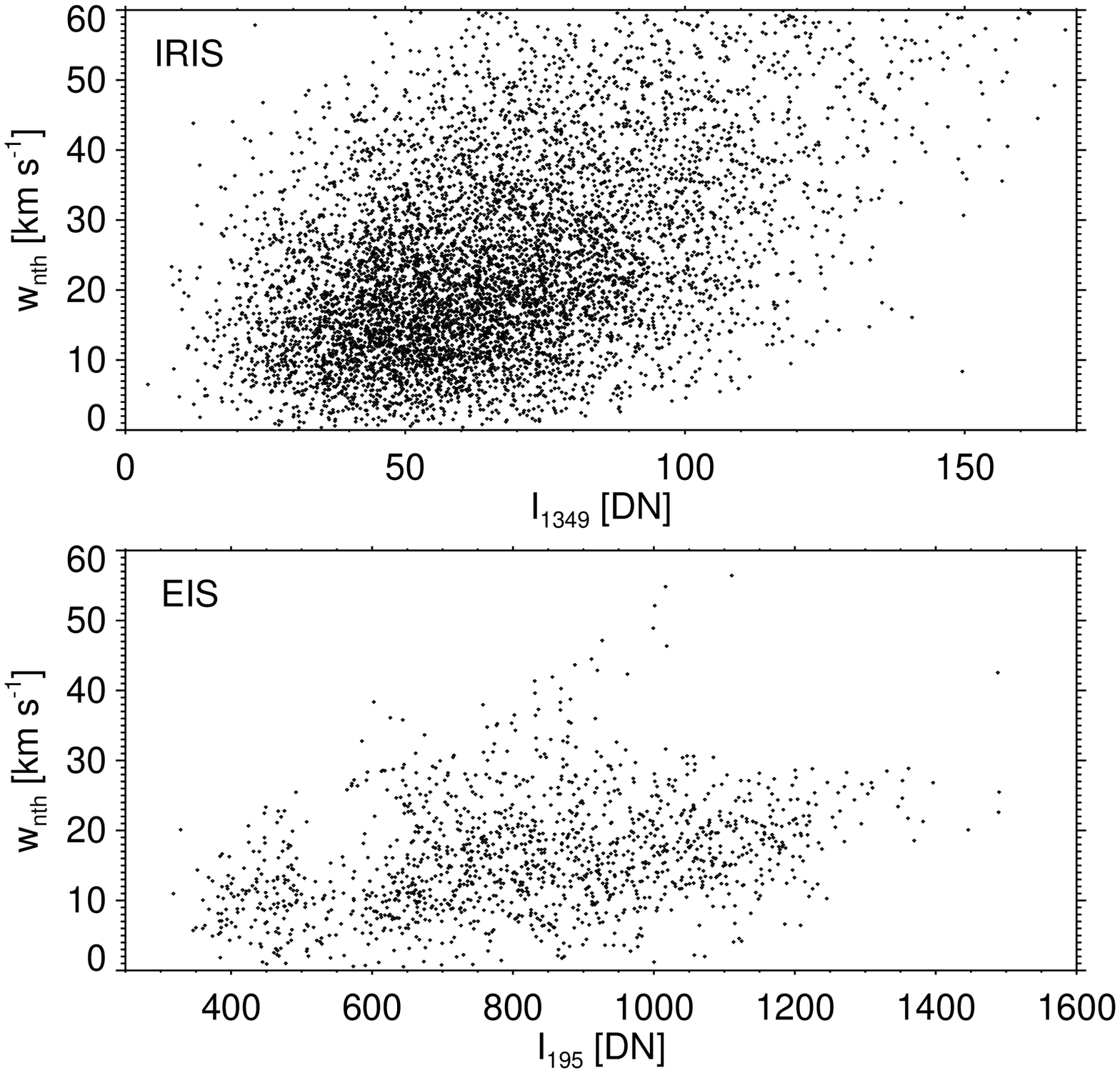}\hspace{-0.8cm}
  \includegraphics[scale=0.65]{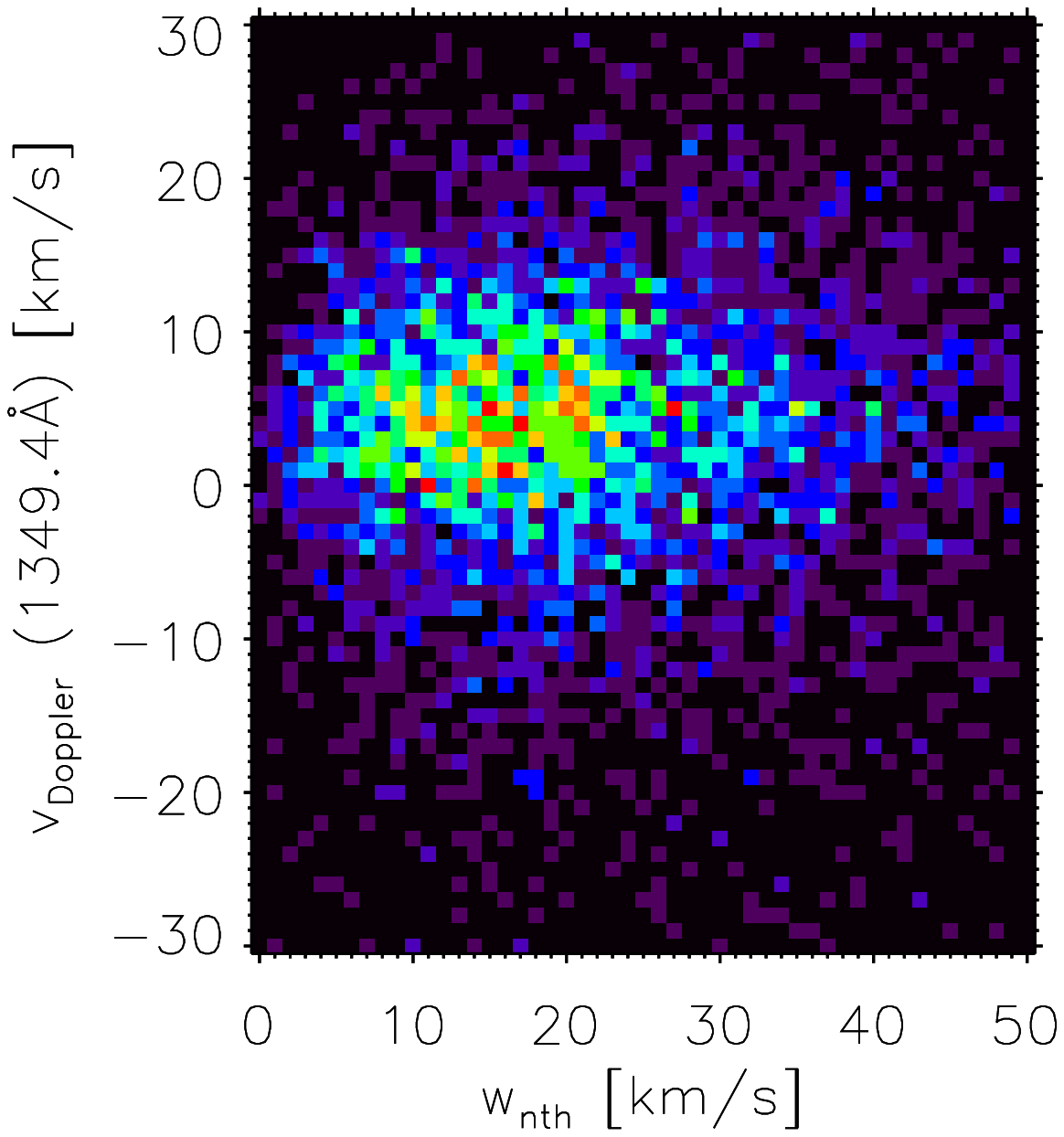}}
\caption{{\em Left:} Correlation between non-thermal line widths
  (corrected for the systematic effects; see Fig.~\ref{fig:fe12_corr}
  and text) and line intensities for the \fexii\ lines in \iris\ (1349\AA; {\em top}) and
  \eis\ (195\AA; {\em bottom}), for the region of overlap of the
  f.o.v.\ of the two spectrographs. 
  {\em Right:} 2D histogram of the \iris\ \fexii\ measured non-thermal
  velocities (x axis) and Doppler shifts (y axis). This plot shows
  an apparent lack of correlation between the two derived quantities.
  For both plots, we selected only the pixels where the \fexii\
  intensity was larger than a threshold value in order to identify the moss.
  \label{fig:correl}} 
\end{figure*}

We investigate the presence of correlations between non-thermal width,
line intensity, and Doppler shifts, which have been found in several
previous spectral studies of active region coronal emission. 
In Figure~\ref{fig:correl} we plot \wnth\ as a function of the line
intensity for both the \iris\ 1349\AA\ line and the \eis\ 195\AA\
line. These plots show the presence of some positive, though weak,
correlation between \wnth\ and line intensity, and that the
correlation is more pronounced at the significantly higher spatial
resolution of \iris. 
Significant fine structuring of the \fexii\ emission on spatial scales
unresolved by \eis\ could explain this result: if \wnth\ is correlated
with line intensity, locally, -- which is plausible as larger non-thermal
velocities could be associated with larger heating, as in the
predictions of several models \citep[e.g.,][]{cargill96,asgari-targhi14}
-- and significant structuring is present on small spatial scale, a
higher spatial resolution would allow to observe more clearly the
correlation between \wnth\ and intensities, which would get washed out
when integrating on spatial scales significantly larger than the
scales of typical structuring.  
The observed behavior of \wnth\ vs.\ line intensity provides clues to
whether noise might be a dominating factor in the determination of
\wnth\ from the \iris\ spectra: if the large \wnth\ values, largely
absent in \eis\ spectra, were attributable to noise, we would expect
some anticorrelation of \wnth\ with intensity, with most of the high
\wnth\ values associated with low line intensities (therefore noisier
spectra); this is not what we observe, therefore the high \wnth\
values are likely real.
The \wnth\ does not appear to have significant correlation with the
Doppler velocity (Figure~\ref{fig:correl}).

\begin{figure}[!ht]
\centerline{\includegraphics[scale=0.5]{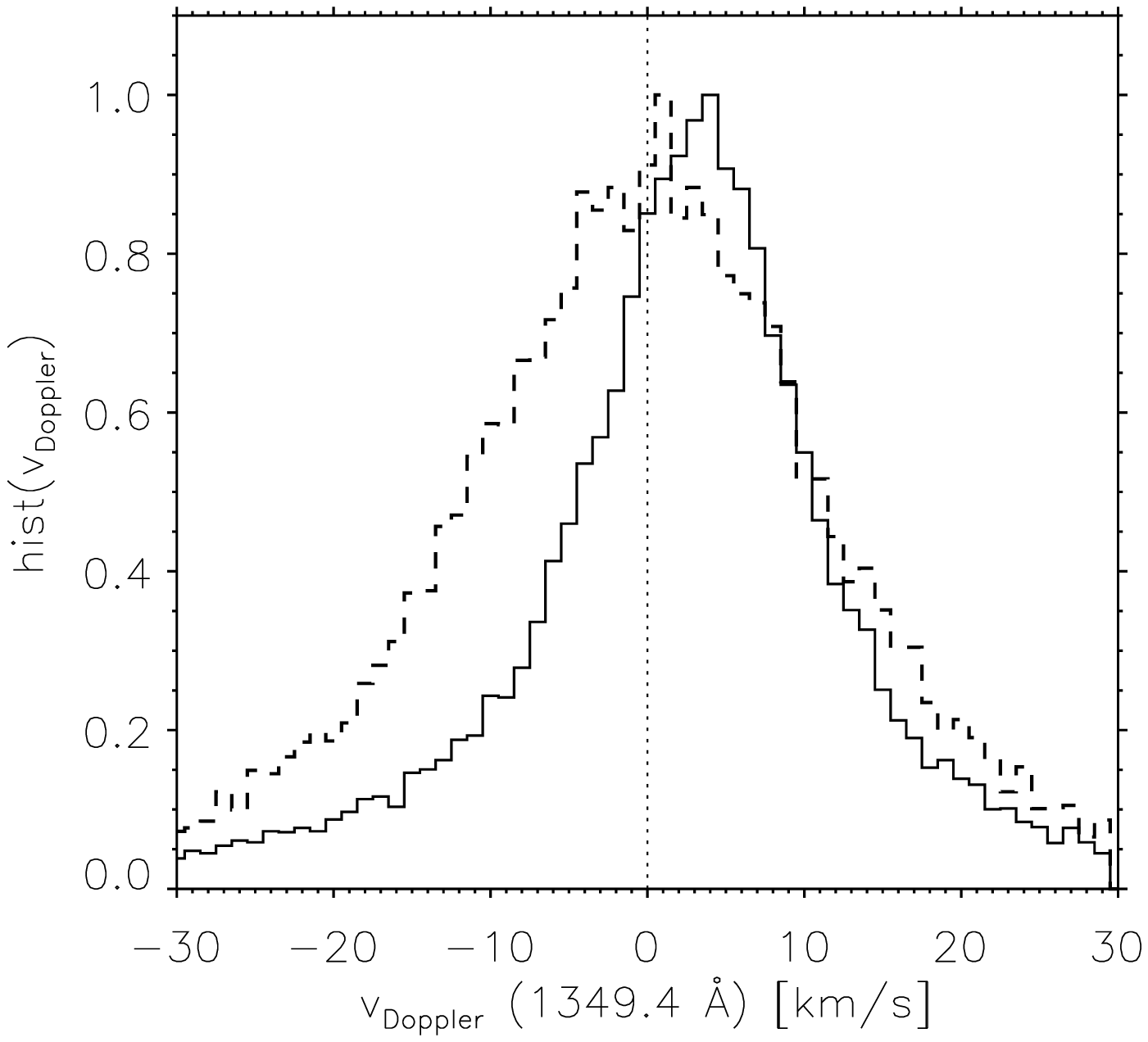}}
\centerline{\includegraphics[scale=0.5]{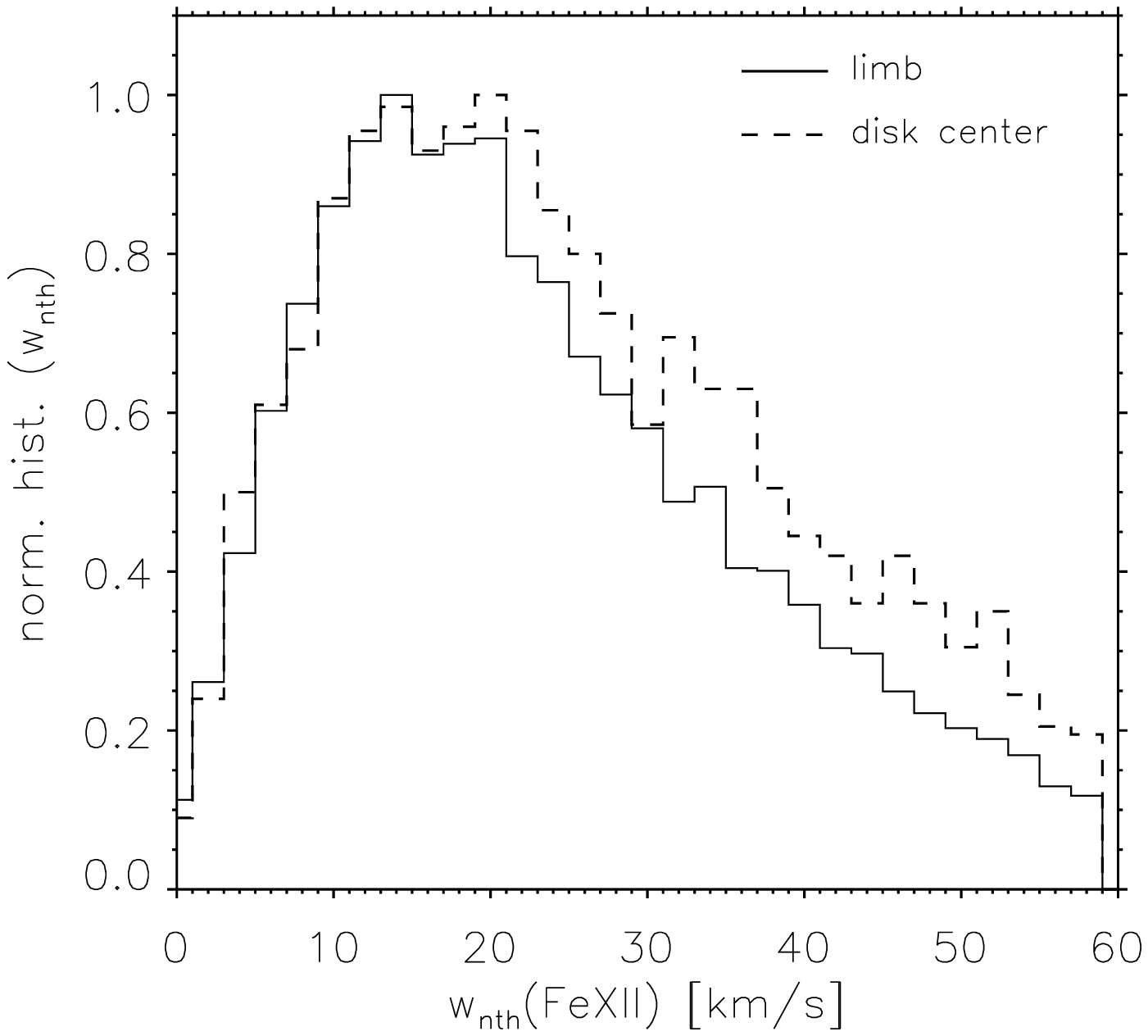}} %\vspace{0.1cm}
\caption{Comparison of distributions of Doppler shifts ({\em top
    panel}) and non-thermal width ({\em bottom panel}) observed with
  \iris\ in moss of AR 12014 at different viewing angles, and at the
  highest spatial resolution ($\sim 0.167$\arcsec$\times
  0.33$\arcsec). We compare the dataset of 2014-03-30, which we
  analyzed in detail in the rest of the paper, with an \iris\
  observation of AR 12014 on 2014-03-26 (00:57UT-01:30UT) when the
  active region was close to disk center (center of \iris\ f.o.v.\ at
  x,y=52.9,-92.8). 
\label{fig:iris_dc_limb}} 
\end{figure}

The distributions of Doppler shifts (see Fig.~\ref{fig:vdoppler}) and
non-thermal line width (Fig.~\ref{fig:fe12_hist}) we found for the
\fexii\ emission observed with \iris\ can have in principle
significant dependencies on the viewing angles. To explore these
effects we consider an additional \iris\ observation of AR 12014 on
2014-03-26 (starting at 00:57UT, i.e., about 4 days prior to the
\iris\ observation of AR 12014 we studied in detail in the rest of the
paper), when the active region was close to disk center (center of
\iris\ f.o.v.\ at x,y=52.9,-92.8), and using the same \iris\ OBSID
(3810263243). Carrying out the same processing and analysis of the
\iris\ spectra we obtain the distributions of Doppler shifts and
non-thermal velocities, and compare them, in
Figure~\ref{fig:iris_dc_limb}, to the results of the observations of
2014-03-30 when the AR is closer to the limb.  We note that any
difference between the two observations at different viewing angle can
in principle be caused by both the viewing angle and by intrinsic
differences of the physical conditions, e.g., due to active region
evolution and different activity level at the two times. The Doppler
shift distribution for the disk center case is broader than for the
case at higher inclination. This could be expected if a significant
portion of the line shifts were due to velocity in a direction close to
the local vertical direction: viewed at a larger angle, the
line-of-sight component of those velocities would become smaller both
on the blue and red side. However, for the observation at high angle
(i.e., when the AR is closer to the limb) the distribution is not only
narrower but also appears redshifted.  This could be interpreted as an
effect of a slightly higher activity level at that time (see discussions in
\S~\ref{s:sims} and \ref{s:conclusions}).
The non-thermal velocity distributions are quite similar to each
other, with the disk center observations showing a larger portion of
pixels at high velocity values (\wnth $\gtrsim 20$km~s$^{-1}$). This
result suggests that field aligned flows significantly contribute to
the observed \wnth, as also suggested by \citet{depontieu15}.

\begin{figure}[!ht]
\centerline{\includegraphics[scale=0.5]{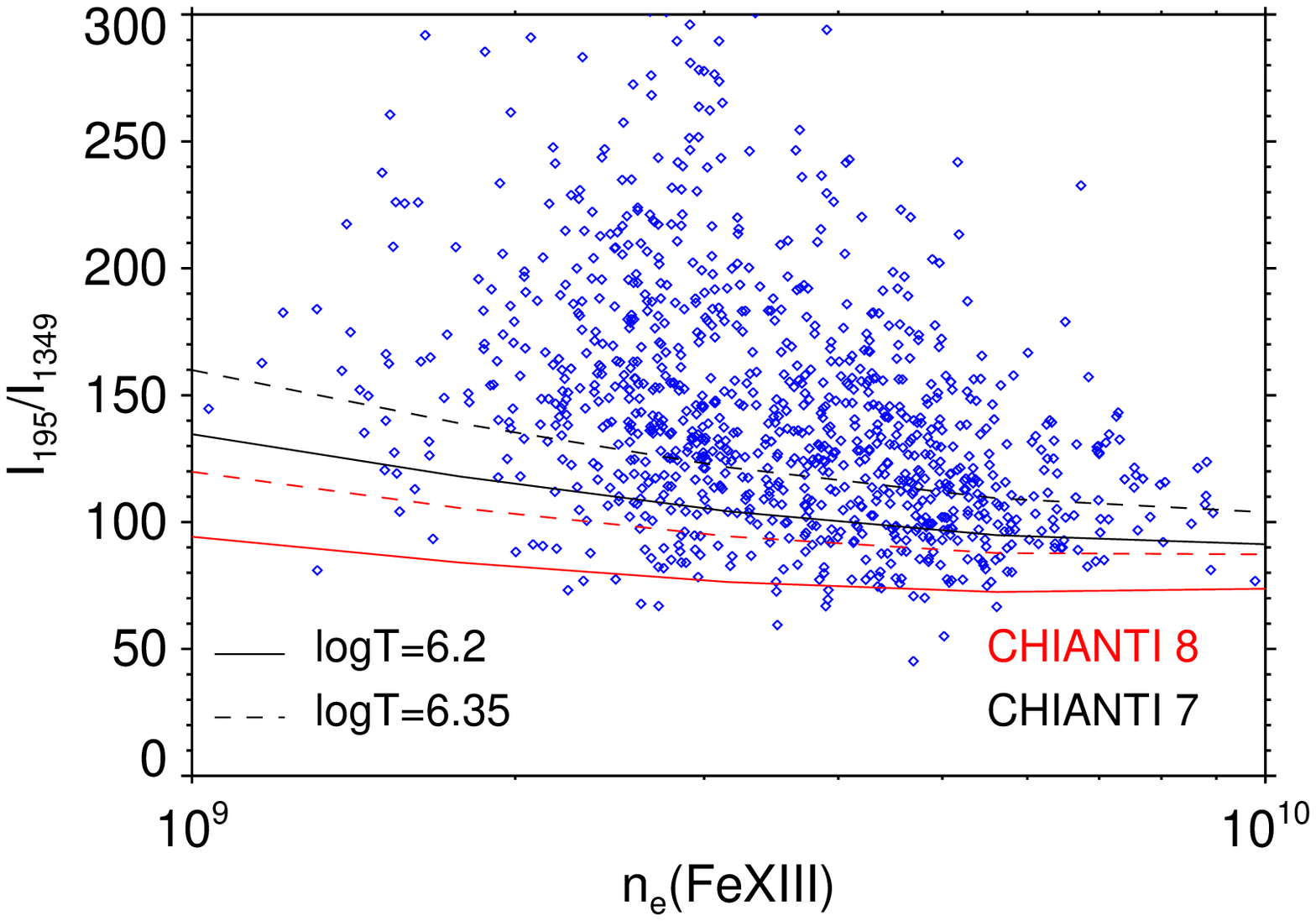}}
\caption{Observed intensity ratio (in
  erg~cm$^{-2}$~s$^{-1}$~sr$^{-1}$) of 195\AA\ \eis\ line and 1349\AA\
  \iris\ \fexii\ line (blue diamond symbols). We show the theoretical
  ratio from the CHIANTI atomic database (version 7, red, and version
  8, black) for two different temperatures ($\log T[K] = 6.2$, solid
  lines, and $\log T[K] = 6.35$, dashed lines). 
  \label{fig:int_ratio}} 
\end{figure}

Finally, we computed the ratios of the measured \fexii\ line
intensities of the \iris\ 1349\AA\ line and the \eis\ 195\AA\ line,
and compared them with the predictions of the CHIANTI database
(Figure~\ref{fig:int_ratio}). In order to calculate this ratio we used
the \fexii\ 1349\AA\ line intensity measured from the \iris\ spectra
rebinned to a 2\arcsec$\times$0.33\arcsec\ spatial scale, and we
converted them to physical units
(erg~cm$^{-2}$~s$^{-1}$~sr$^{-1}$). The uncertainties in the effective
areas of \iris\ and \eis\ are expected to be of the order of 20-25\%
(\citealt{depontieu14,lang06,delzanna13,warren14}). 
We calculated the CHIANTI predicted line ratios using the last two
CHIANTI versions (in the just released v.8 of CHIANTI, some of the
\fexii\ data in the EUV wavelength range have significantly
changed), and for two different temperatures. 
The comparison between the observed and the theoretical values shows
that the atomic data systematically underestimate the 195\AA/1349\AA\
ratio (i.e., they predict a stronger than observed 1349\AA\ line,
compared with the 195\AA\ line), and that the discrepancy is larger
for CHIANTI 8. We note that in reality the extent of
the underestimate is likely even larger, because the 195\AA\ emission
in moss is expected to suffer significant (typically a factor $\sim
2$) absorption from cool chromospheric material, which affects
wavelengths below 912\AA, and it is due to resonance continua of
neutral hydrogen and helium \citep{depontieu09}.

\section{3D radiative MHD simulations}
\label{s:sims}

In order to gain insight into the observed \fexii\ spectral properties
of active region moss we consider `realistic' 3D radiative MHD
simulations of the solar atmosphere using the Bifrost code
\citep{Gudiksen11}. The Bifrost code solves the full resistive MHD equations,
including non-LTE and non-grey radiative transfer with scattering, and with
thermal conduction along the magnetic field lines. 
The model spans from the upper layer of the convection zone up to the
low corona, and self-consistently produces a chromospheric and hot
corona, through the Joule dissipation of electrical currents that
arise as a result of foot point braiding in the photosphere and
convection zone \citep{hansteen15}.

\begin{figure}[!ht]
\centerline{\includegraphics[scale=0.7]{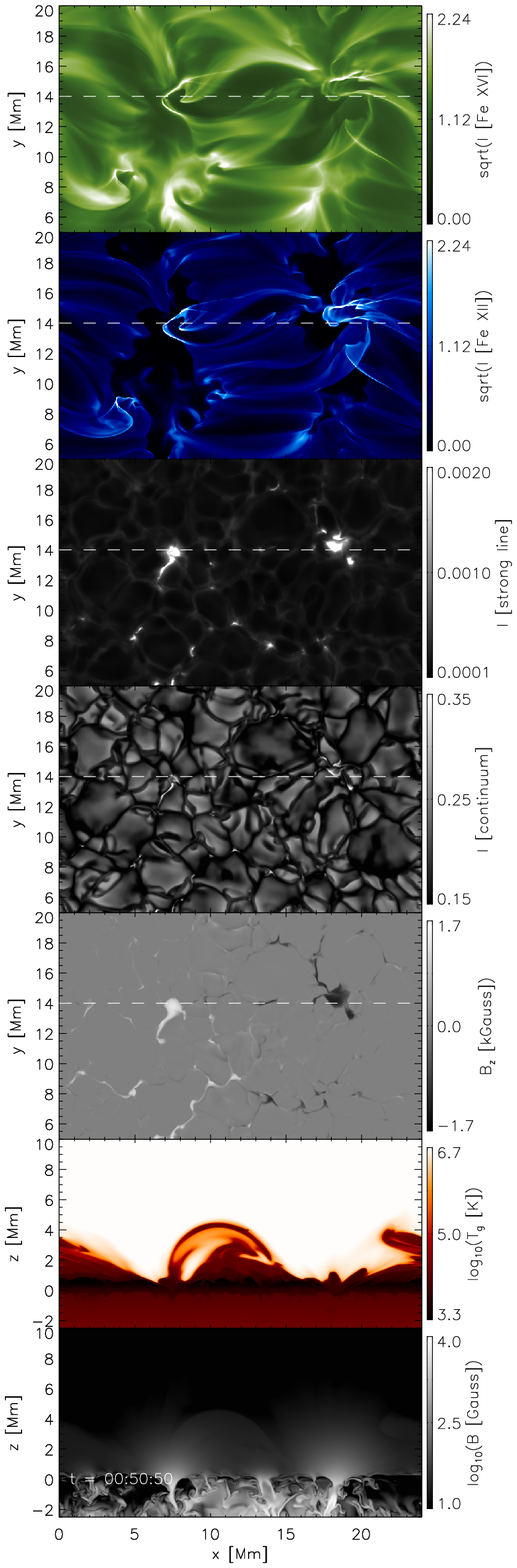}}
\caption{Spatial distribution of several physical variables in the
  Bifrost simulation we analyze here. From {\em top} to {\em bottom}:
  (a) top view \fexvi\ (335\AA) intensity showing the structure of the
  $\sim 2-3$~MK corona; (b) top view of \fexii\ (195\AA) intensity
  showing that the highest intensity regions consist of moss-like
  areas at the footpoints of hot loops; (c) top view line intensity of
  bin4 intensity (see text) with opacity typical of a strong
  chromospheric line; (d) top view intensity of the bin1 (low opacity,
  see text) emission typical of the solar continuum emission in the
  visible; (e) top view intensity of vertical component of magnetic
  field; (f) side view of plasma temperature; (g) side view of
  magnetic field intensity. The white dashed line in the top five
  panels shows the location of the vertical cut displayed in the
  bottom two panels. 
  \label{fig:bifrost_hot}} 
\end{figure}
 
Here we consider a Bifrost simulation yielding a high temperature
corona ($\gtrsim 5$~MK) and therefore having strong \fexii\ emission
closer to the loop footpoints (i.e., ``moss''; see
Figure~\ref{fig:bifrost_hot}). 
The simulation covers a region of dimensions $24 \times 24 \times
17$~Mm$^3$, with $768 \times 768 \times 768$ grid points ($\Delta x =
\Delta y \sim 31$~km, $\Delta z \approx 13$~km up to a height of 5~Mm
above the photosphere  and increasing to $\approx 80$~km at the top of
the computational box).  

The magnetic field configuration of this model is very similar to that
found in the publicly available simulation published as part of the \iris\ project
and described in \cite{carlsson16}. In both models the
photospheric field is dominated by two concentrated opposite polarity
regions of approximately equal strength that are some 10~Mm
apart. This configuration gives a set of loops in the chromosphere and corona
connecting the opposite polarity regions. In addition (and as opposed
to the publicly available model) a weaker, 100~G, horizontal field is
continuously injected at the bottom boundary, 2.5~Mm below the
photosphere. This injection eventually leads to a weaker `salt and pepper'
field that fills the entire photosphere at the time of the model snapshots
presented here.
The additional magnetic field alters the atmospheric conditions
sufficiently to raise the coronal temperature well above what is found
in the publicly available model. 
It is the upwardly directed Poynting flux, generated by the interaction
of the photospheric motions with the magnetic field, that ultimately
heats the outer layers of the model to high temperatures.

Figure~\ref{fig:bifrost_hot} shows the spatial distribution of several
physical variables for a snapshot of the Bifrost simulation near the
time of largest coronal temperatures and analyzed
here; we show side views of the total magnetic field and temperature and top 
views of the vertical component, $B_z$, of the photospheric field, 
the continuum, chromospheric and transition region and coronal
intensities. 
The optically thick radiative losses are computed by binning solar
opacities into four bins, sorted according to magnitude, as described
by \cite{nordlund82,skartlien00,hayek10}. In Figure
~\ref{fig:bifrost_hot} we show the emission in bin 1 (similar to
  white light emission with an effective temperature of approximately
  5800K), which represents
the lowest opacities, and bin 4 which represents the highest opacities
of those considered. Bin 1 is similar to continuum emission at visible
wavelengths in the solar spectrum, while bin 4 is typical of a strong
chromspheric line such as \caii\ H or K. 
The strong magnetic field concentrated in the 
photosphere rapidly spreads in the chromosphere and forms loop-like
structures connecting the opposite polarity regions that penetrate
into the corona. High temperature coronal material penetrates to
within a 1~Mm of the photosphere in the vicinity of the footpoints of
these loops. The synthetic images in the \fexvi\ and \fexii\ emission show
that the simulated atmosphere has hot loop-like coronal structures,
and the highest \fexii\ emission is concentrated at the footpoint of
hot, dense loops. Therefore we can use the synthetic \fexii\ emission as a
comparison with our moss observations.

\begin{figure*}[!ht]
\centerline{\includegraphics[scale=1.2]{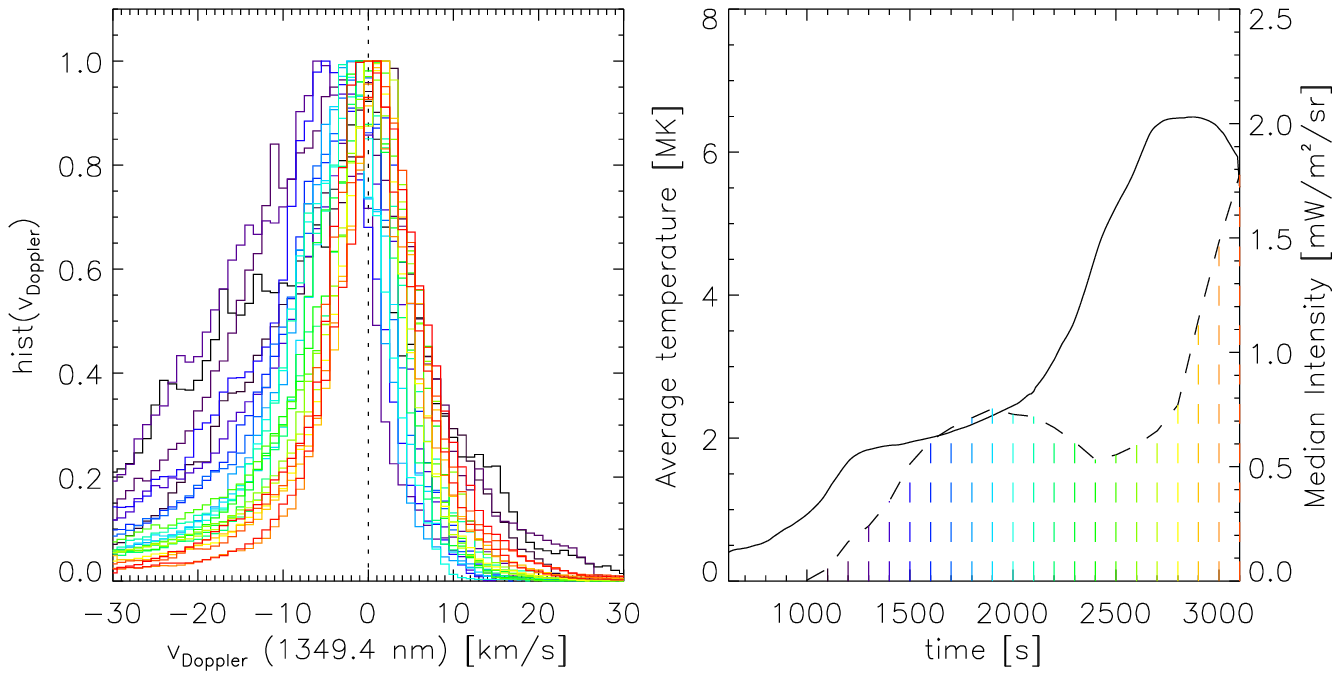}}
\caption{{\em Left:} Histograms of Doppler velocity distributions of
  \fexii\ emission in the Bifrost 3D MHD simulation at different times
  in the evolution of the atmosphere (see {\em right panel} for the
  correspondence between time and color). {\em Right:} Evolution of
  temperature (solid line) and \fexii\ median intensity (dashed line)
  in the Bifrost simulation.
  \label{fig:sim_v_hist}} 
\end{figure*}

From the Bifrost simulation we calculate the Doppler shift and the
non-thermal width of the 1349\AA\ \iris\ line, to compare it with our
\iris\ observations. In the \iris\ observation, the sensitivity of
the instrument effectively selects only the brightest \fexii\ emission
regions, i.e., the moss. 
For a more meaningful comparison with the \iris\ observations we
calculate the distributions of the \fexii\ spectral properties on a
subset of the simulation volume, where the \fexii\ is highest (i.e.,
at the footpoints of hot loops), using $3 \times$ the median
intensity as a threshold value. We also integrate the \fexii\ emission
on the same spatial resolution as with \iris, and integrate over 30~s; note that
we have one snap shot every 10~s, therefore we may
underestimate the range of dynamics that is present in both the model
and the real solar atmosphere on much smaller temporal scales.  

In Figure~\ref{fig:sim_v_hist} we show the histograms of the 1349\AA\
\fexii\ Doppler shift for a series of snapshots (one every 100~s) in
the interval 1000-3000~s from the start of the simulation. 
As the atmosphere is heated to high coronal temperatures (the average
temperature in the simulation box is plotted in the {\it right panel}
of Figure~\ref{fig:sim_v_hist}), the peak of the Doppler shift
distribution goes from blueshifts (upflow) of 5-10~km~s$^{-1}$, via
smaller blueshifts, and finally to low amplitude redshifts (downflow)
of some few km~s$^{-1}$.  
This can be explained by the fact that \fexii\ is shifting from
being emitted mainly from coronal plasma, when the corona is
relatively cool ($\lesssim 2$~MK), to being emitted from transition
region plasma, confined by a large temperature gradient, when the
corona gets hotter.  
As the coronal temperature increases, the plasma most strongly
emitting \fexii\ lines (i.e., plasma typically at 1-1.5~MK) becomes more
and more confined to the lower layers of coronal structures (i.e., at
the loop footpoints) and therefore the observed \fexii\ lines behave
more like typical transition region lines, generally dominated by
redshifts. 
Both \cite{peter06} using the Stagger code, and \cite{Hansteen10}
using Bifrost simulations found that the observed net transition
region redshifts were reproduced in 3D 'realistic' numerical models.
As described in \cite{Hansteen10}, in the Bifrost simulations, 
redshifts of the transition region emission are a consequence of the
episodic heating and concentration of the heating per particle in the
lower atmosphere: the plasma is heated at lower heights to high
temperature, and the increased pressure causes upflows in the hotter
lines and downflows in the cooler (transition region) lines.
The evolution of median intensity of \fexii\ emission as a function of
time in the simulation is also shown in the {\em right panel} of
Figure~\ref{fig:sim_v_hist}. The \fexii\ intensity initially increases
as the coronal temperature increases from initial values of $\sim
0.5$~MK to $\sim 2$~MK. As the coronal temperature continues to 
increase, the \fexii\ emission initially decreases as less and less
of the coronal volume emits \fexii\ since it becomes too hot. 
Later ($t > 2500$~s) the emission increases again, and steeply, 
as the \fexii\ emission is ``pushed'' to steadily denser transition 
region layers.

\begin{figure}[!ht]
\centerline{\includegraphics[scale=1.2]{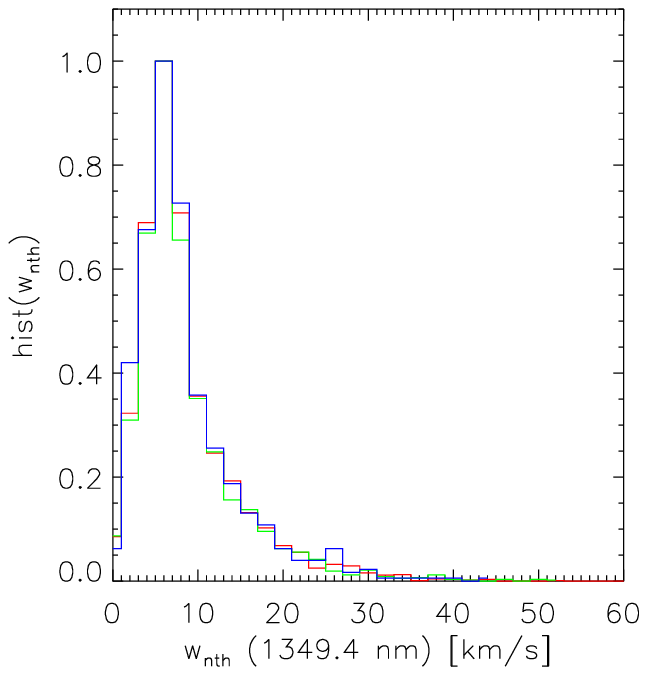}}
\caption{For the Bifrost snapshot at time t=3050~s, when the coronal
  temperature is close to its peak (see Figure~\ref{fig:sim_v_hist}), we
  show the histograms of \fexii\ non-thermal line width distributions
  for the same spatial binning used for the analysis of the \iris\
  data, with the same color coding of Figure~\ref{fig:fe12_hist}: 
  red for the \iris\ original spatial resolution
  ($0.167$\arcsec$\times 0.33$\arcsec), green for a rebin of a factor
  4 ($0.66$\arcsec$\times 0.33$\arcsec), and blue for a rebin of a
  factor 12 ($2$\arcsec$\times 0.33$\arcsec).
  \label{fig:sim_w_hist}} 
\end{figure}
 
In Figure~\ref{fig:sim_w_hist} we plot the histogram of the
non-thermal line widths of the \fexii\ 1349\AA\ emission from the
Bifrost simulation (for a snapshot where the average coronal
temperature is $\sim 6$~MK). The spectral line profiles are
synthesized from the Bifrost snapshot by using the \iris\ spectral,
spatial and temporal resolution, and selecting only the strongest
emission regions (as described above). After \iris-like spectra are
obtained, they have been analyzed at three different levels of spatial
binning, just like for the actual data (see Figure~\ref{fig:fe12_hist}). 
The non-thermal line widths, \wnth, are then calculated as for the
actual spectra by subtracting instrumental and thermal broadening (see
beginning of \S~\ref{s:results}). The \wnth\ distributions from the
Bifrost simulation peak at lower values ($\sim 7$~km~s$^{-1}$)
than for the observations ($\sim 15$~km~s$^{-1}$, see
Figure~\ref{fig:fe12_hist}), and they are also characterized by
narrower distributions.  
The limited non-thermal broadening from Bifrost simulations compared
with observations has been observed and discussed in previous papers
\cite{Hansteen10,olluri15}. 
The effect of decreased spatial resolution (i.e., larger spatial bin)
in the simulated spectra is similar but not identical to that seen for
the actual data: at lower spatial resolution the peak of the \wnth\
distribution does not change, but in contrast with the data, we do not
see any narrowing of the distribution with higher spatial resolution,
except in certain subsets.

\section{Discussion and conclusions}
\label{s:conclusions}
We have presented a first analysis of \iris\ \fexii\ spectral
observations at the highest spatial resolution to date, which show the
presence of fine coronal structure on subarcsecond scale. In
particular we have shown that \fexii\ emission can be observed with
\iris\ in bright (dense) coronal structures such as post-flare loops,
and active region moss, by adopting long exposure times, and lossless
compression.

The presence of neutral lines in the \iris\ spectra allow an accurate
absolute wavelength calibration, and therefore a more
accurate determination of line Doppler shifts, compared to some other
spectral observations, e.g., with \eis.
For the post-flare loops we find that the \fexii\ emission is largely
unshifted at the loop tops and predominantly redshifted at both loop
footpoints, as expected for draining of plasma in the later phases of
the flare when the heating has ceased.
For the moss observations, we find that the distribution of Doppler
shifts is peaked at a redshift of a few km~s$^{-1}$, but with
significant wings on both the red and the blue side.  
In our interpretation, guided by the Bifrost 3D models, \fexii\
appears redshifted for hotter coronal temperatures ($\gtrsim 4$MK),
while blueshifted when the corona is at cooler temperatures. The
distribution of Doppler shift we observe shows a wide range of blue
and red shifts, and it suggests that there may be a continuous mix of
heating and cooling in the moss we observed. 
For cooler transition region lines ($\log T$[K]$\sim 4.7-5.7$) the
dominance of redshift is well established
\citep[e.g.,][]{doschek76,dere84,achour95,chae98b,peter99}. 
The physical processes causing the observed redshifts are still
debated, and several models have been proposed
\citep[e.g.,][]{hansteen93,Patsourakos06,peter06,Hansteen10}.
\cite{Hansteen10} use 3D radiative MHD models of the solar atmosphere
to investigate the transition region redshifts, and find that in their
model the heating is dominated by rapid intermittent events at low
heights which heat the plasma locally to coronal temperature and
produces downflows at transition region temperature due to the local
overpressure. The model analyzed by \cite{Hansteen10} reached modest
coronal temperatures ($\lesssim 2.5$~MK), therefore with significant
\fexii\ emission in loop structures rather than at the loop footpoints
as in the case for moss. Here we used a different Bifrost 3D MHD
simulation which reaches higher coronal temperature and represent a
better model for moss observations.
We find that the moss \fexii\ (mostly) redshifted emission we observe
with \iris\ is reproduced in the model when the corona gets hot enough
and most of the \fexii\ is confined in the loops transition
region. The \fexii\ then behaves like a transition region line, and is
redshifted, in the model, as a consequence of the local heating at low
heights which causes upflows in the hotter lines and downflows in the
cooler (transition region) lines. 

We also analyzed the \fexii\ line broadening, in both \iris\ and \eis\
moss observations, and find that the non-thermal broadening is
typically small (the distributions peak around 15~km~s$^{-1}$) and the
\wnth\ distributions are generally similar at different spatial
resolution scales, as found also by \cite{depontieu15} for the cooler
\iris\ \siiv\ emission. However, we find that at the higher spatial
resolution accessible with \iris\ the \wnth\ histograms are broader.
The fact that at higher resolution, higher values of \wnth\ are
observed, is somewhat counterintuitive, if interpreting the \wnth\
as due to unresolved motions along the line-of-sight. The higher
\wnth\ observed at higher spatial resolution must come from
spatially separate events, occurring on a spatial scale typically
smaller than \iris\ spatial resolution, and with intrinsic large
mixing of velocities along the line-of-sight. These features are
possibly associated to heating events. 

We discussed in detail systematic effects affecting the line
broadening measurements in both \iris\ and \eis: low signal-to-noise
ratio leads to systematic underestimates of \wnth\ from \iris\
spectra, while the absolute calibration of \eis\ causes systematic
overestimates of \wnth\ (see \citealt{brooks15}). These effects need
to be taken into account for a correct measurement and interpretation
of the spectral lines broadening. 
We analyzed the non-thermal broadening in the Bifrost 3D model and
find that the \wnth\ are systematically smaller than in the
observations (the distributions peak around 10~km~s$^{-1}$) but
otherwise they are affected by spatial resolution in a similar way to
the observations. We note however that the comparison of the
modeled and observed velocities might be significantly affected by
the viewing angle.
The low average \wnth\ that we find from the \fexii\ spectral analysis
are in agreement with several previous findings (e.g.,
\citealt{dere93,Brooks09}, though at coarser spatial resolution), but
at the \iris\ subarcsecond resolution we also find a significant tail
at larger \wnth\ values, which can further constrain the models. 
In particular, the high \wnth\ values might be missing in current
  models because of insufficient spatial resolution, as test
  simulations at higher resolution show a significant increase in
  turbulence, vorticity and small-scale flows in the solar atmosphere
  that could lead to significant non-thermal broadening in the
  atmosphere. In addition, preliminary results suggest that the mix of
  strong fields and neighboring mixed polarity weaker fields plays an
  important role in creating atmospheric dynamics. Therefore,
  simulations with different photospheric magnetic field distributions
  could well lead to an increase of non-thermal line broadening.
\cite{brooks15} recently investigated the non-thermal broadening of
hot lines (up to $\sim 5$~MK; see also \citealt{imada09}) in the core
of active regions and find that they are very modest ($\lesssim
15$~km~s$^{-1}$; though see also \citealt{saba91}).  
The hot line measurements by \cite{brooks15} are
made in the coronal portion of hot loops at the footpoints of which is
the moss. Our \wnth\ results together with \cite{brooks15}, assuming
that both are typical of active regions, suggest that the \wnth\ does
not increase with temperature, in disagreement with some models
\citep[e.g.,][]{Patsourakos06}. 
As discussed in \S~\ref{s:results} and here above, the measured
velocities (both Doppler shifts and non-thermal velocities) will be
affected by the viewing angle. We investigated the possible viewing
angle effects by comparing \iris\ \fexii\ observations of the same
active region moss for significantly different angles. The
observations we considered were taken 4 days apart therefore some
differences are likely due to active region evolution. However,
the fact that the \wnth\ distribution is broader for the near disk
center observation, compared to the near limb observation, seem to
indicate that the field aligned velocities provide a significant
contribution to the \wnth. In a scenario in which the \wnth\ were
largely due to velocities perpendicular to the field lines (e.g.,
for Alfv$\acute{\rm e}$n wave turbulence models; e.g.,
\citealt{vanballegooijen11,vanballegooijen14}) we would expect
smaller \wnth\ for observations near disk center, for which our line
of sight is expected to be more parallel to field lines on average.

Previous spectral studies of active region coronal emission have
explored correlations between non-thermal width, line intensity, and
Doppler shifts. In our analysis we find for \fexii\ no correlation
between Doppler shifts and non-thermal widths, and a weak correlation
between \wnth\ and line intensities.
Early studies with \hrts\ focused on cooler transition region
lines (e.g., \civ, \siiv), and found some correlation between \wnth\ and
line intensities \citep[e.g.,][]{dere84,dere93}. 
\sumer\ analyses expanded on these early findings, however, mostly
focusing on quiet sun observations as only few on-disk active region
\sumer\ observations are available to compare with our observations. 
Quiet sun \sumer\ studies have generally found some level of
correlation between \wnth\ and line intensity, e.g., in \oiv, \ov,
\nv, \siiv\ \citep[e.g.,][]{warren97,landi00,akiyama05}. 
A comprehensive QS study by \cite{chae98} explored the \wnth-intensity
correlation for a broad range of temperatures and found that the two
parameters are correlated for lines emitted at temperatures in the $2
\times 10^{4}-10^5$~K range, but the correlation becomes significantly
weaker for hotter lines.
\sumer\ observations including active regions show some cases where
the \wnth-intensity correlation is not very clear (possibly because of
the mixing of different types of solar coronal features; e.g.,
\citealt{spadaro00}), and some cases where \wnth\ appears correlated
with intensity for cooler lines (up to \civ) but not for the hotter
\neviii\ line (\citealt{feldman11}, echoing the QS results of
\citealt{chae98}). 
More recently, \hinode-\eis\ studies have analyzed the possible
correlations of \wnth\ and line intensity for a wide range of
spectral lines (and plasma temperatures), and coronal features.
The correlation of \wnth\ and Doppler velocity is mostly found in
outflow regions at the edges of active regions
\citep[e.g.,][]{doschek08}, while the two are not correlated in moss
\citep{doschek12}. 
\cite{li09} and \cite{scott11} investigate the correlations between
\wnth\ and line intensities and find contrasting results: positive
correlation \citep{li09} or negative correlation \citep{scott11}.
We argue that different studies include varied combinations of solar
features, significantly affecting the presence of correlations. Also,
these cited previous studies do not specifically address moss emission
and therefore make for a poor comparison with our results.

In conclusion, in this work we analyzed spectral observations of
  \fexii\ emission, for the first time at subarcsecond resolution
  accessible with \iris, in particular focusing on the Doppler shifts
  and non-thermal widths of the emission of active region moss.
  We find that \fexii\ moss emission is mildy redshifted on average, and
  shows broad distributions of Doppler shifts. These characteristics
  are reproduced by a 3D MHD Bifrost simulation with a hot corona
  ($\gtrsim 4$~MK), in which the \fexii\ emission is largely confined
  to the transition region and shows redshifts caused by episodic
  heating in the low solar atmosphere.
  For the non-thermal line width distribution we find that the peak
  does not appear to depend on spatial resolution, but with increasing
  resolution the distributions become broader with both more low and
  high values of non-thermal line broadening. Analogous results have
  been found for \siiv\ emission observed with \iris\
  \citep{depontieu15}.  
  If the tail of high \wnth\ values is due to heating events, it
  suggests that these events typically happen on small (subarcsecond)
  spatial scales, and their temporal frequency is such that not many
  of them happen simultaneously on arcsecond scale areas.
  Finally, our observations of the same active region from different
  viewing angles show a slight increase in non-thermal broadening at
  disk center suggesting that field-aligned flows contribute
  significantly to the broadening. Our results provide constraints for
  heating models based on Alfvenic turbulence which predict increased
  broadening for viewing angles perpendicular to the magnetic field. 

\begin{acknowledgements}
We thank P.\ Young for providing help with fitting routines for the
\eis\ and \iris\ analysis, and for providing the \fexii\ CHIANTI v.8
data before their public release. We thank D.\ Brooks for useful
discussions about analysis of non-thermal line widths from
\hinode-\eis\ spectra. We also thank the anonymous referee for
insightful comments which helped us to improve the paper.
PT was supported by contract 8100002705 from Lockheed-Martin to SAO,
NASA contract NNM07AB07C to the Smithsonian Astrophysical Observatory,
and NASA grants NNX15AF50G and NNX15AF47G. 
The authors thank the International Space 
Science Institute (ISSI) for support to the teams ``New Diagnostics of Particle
Acceleration in Solar Coronal Nanoflares from Chromospheric
Observations and Modeling'', and ``Heating in the magnetic
chromosphere'', where topics relevant to this work were discussed with
other colleagues.  
\hinode\ is a Japanese mission developed and launched by ISAS/JAXA, with 
NAOJ as domestic partner and NASA and STFC (UK) as international 
partners. It is operated by these agencies in co-operation with ESA 
and the NSC (Norway).
\iris\ is a NASA small explorer mission developed and operated by LMSAL
with mission operations executed at NASA Ames Research center and
major contributions to downlink communications funded by ESA and the
Norwegian Space Centre.
\end{acknowledgements}

\bibliographystyle{/Users/ptesta/WORK/PAPERS/apj}

\end{document}